\documentclass{ar-1col}
\usepackage{url}
\usepackage[numbers]{natbib}
\usepackage[letterpaper, portrait, margin=1.5in]{geometry}
\usepackage{amsmath}
\usepackage{mathrsfs}
\usepackage{bm}
\usepackage[cmintegrals]{newtxmath}
\usepackage{multirow}
\usepackage{natbib}
\usepackage{graphicx}
\usepackage{longtable}
\usepackage{tabularx}
\usepackage{lineno}
\usepackage[colorlinks,linkcolor=blue,citecolor=blue,urlcolor=blue]{hyperref}

\jname{Xxxx. Xxx. Xxx. Xxx.}
\jvol{AA}
\jyear{YYYY}
\doi{10.1146/((please add article doi))}

\newcommand{\bk}{{\bm k}}

\newcommand{\etal}{{\it et al.}}
\newcommand{\ie}{{\it i.e.~}}
\newcommand{\eg}{{\it e.g.~}}

\begin{document}


\markboth{Gao {\etal} et al.}{Short title}

\title{Topological Semimetals from first-principles}

\author{Heng Gao,$^1$ J\"orn W. F. Venderbos,$^{2,3}$ Youngkuk Kim,$^4$ and Andrew M. Rappe$^2$
\affil{$^1$International Centre for Quantum and Molecular Structures, Department of Physics, Shanghai University, 99 Shangda Road, Shanghai 200444, China}
\affil{$^2$Department of Chemistry, University of Pennsylvania, Philadelphia, Pennsylvania 19104--6323, USA}
\affil{$^3$Department of Physics and Astronomy, University of Pennsylvania, Philadelphia, Pennsylvania 19104--6396, USA}
\affil{$^4$Department of Physics, Sungkyunkwan University, Suwon 440-746, Korea}}

\begin{abstract}
We review recent theoretical progress in the understanding and prediction of novel topological semimetals. Topological semimetals define a class of gapless electronic phases exhibiting topologically stable crossings of energy bands. Different types of topological semimetals can be distinguished based on the degeneracy of the band crossings, their codimension (\eg point or line nodes), as well as the crystal space group symmetries on which the protection of stable band crossings relies. The dispersion near the band crossing is a further discriminating characteristic. These properties give rise to a wide range of distinct semimetal phases such as Dirac or Weyl semimetals, point or line node semimetals, and type-I or type-II semimetals. 
In this review we give a general description of various families of topological semimetals with an emphasis on proposed material realizations from first-principles calculations. The conceptual framework for studying topological gapless electronic phases is reviewed, with a particular focus on the symmetry requirements of energy band crossings, and the relation between the different families of topological semimetals is elucidated. In addition to the paradigmatic Dirac and Weyl semimetals, we pay particular attention to more recent examples of topological semimetals, which include nodal line semimetals, multifold fermion semimetals, triple-point semimetals.
Less emphasis is placed on their surface state properties, responses to external probes, and recent experimental developments. 
\end{abstract}

\maketitle


\section{INTRODUCTION \label{sec:intro}}

The discovery of topological insulators (TIs) firmly established the notion of topology as a means to sharply distinguish electronic phases and classify quantum states of matter~\cite{Hasan10p3045, Qi11p1057}. Classification schemes based on the topology of the ground state wave function---initially applied to electronic phases with an energy gap---have since been extended to gapless systems~\cite{Armitage18p015001}. This has led to the identification of a new and special class of metals: the topological semimetals (TSMs). TSMs are characterized by a topologically stable Fermi surface originating from a crossing of energy bands. Band crossings of this kind can be associated with a topological number, which may depend on the symmetries responsible for enforcing or protecting the band crossing degeneracy. 

Different types of TSMs can be distinguished based on key attributes of the band crossing, such as its degeneracy, the codimension (\ie whether the band degeneracy occurs at a point or on a line), and the dispersion in the vicinity of the crossing. Another possible distinction can be made based on the origin of the crossing, that is to say, whether it is symmetry-enforced or arises as a result of a band inversion. These attributes, in combination with their topological characteristics, have led to the identification of a growing number of different TSM families, which include Dirac and Weyl semimetals~\cite{Young12p140405,Wang12p195320,Wan11p205101,Weng15p011029}, nodal line semimetals~\cite{Horava05p016405, Heikkia11p59, Burkov11p235126}, type-I and type-II semimetals~\cite{Soluyanov15p495}, multifold fermion semimetals~\cite{Wieder16p186402,Bradlyn16p5037}, and ``triple-point'' semimetals~\cite{Zhu16p031003, Weng16p241202, Chang17p1688, Wieder18p329}, among others. 

Topological semimetals can host a variety of different low-energy excitations, which not only offer promising potential for future applications but also offer a new platform for the fundamental study of novel quasiparticles beyond the standard paradigm of known particles in high-energy physics~\cite{Bradlyn16p5037}.  Due to the nontrivial topology of the bulk and surface electronic states, topological semimetals are expected to exhibit intriguing quantum transport properties, such as unusual magnetoresistance and the chiral anomaly~\cite{Jeon14p851,Feng15p081306,Zhang15p02630,Huang15p031023,Wang16p13142}, which have attracted broad attention from both theoretical and experimental communities. Furthermore, topological semimetals are of interest due to their potential future application in chemical catalysis~\cite{Rajamathi17p1606202, Li18p23}, quantum computation~\cite{Nayak08p1083}, and spintronics~\cite{Yang16p1640003}.

In this review we present an overview of the recent progress---mainly from the theoretical frontier---in understanding and predicting novel types of TSMs, with an emphasis on first-principles electronic structure calculations. In the past few years, this field has witnessed a rapid development, ranging from the advancement of conceptual ideas to specific predictions of new materials. First-principles calculations have played an important role in this development by providing guidance and support in the search for new topological materials in general and TSMs in particular. A further aim of this review is to convey the central pillars on which the theory of topological semimetallic phases rests, paying particular attention to the connections and relationships that exist between different families of TSMs. A particularly useful theme which makes these connections transparent is the notion of symmetry. The symmetry properties of solid state materials are determined by their crystal and compositional structure and in turn determine the degeneracies and crossings of energy bands. Symmetry therefore naturally relates structural and electronic characteristics. As a result, the symmetries of the crystal lattice, \ie the symmetries of the space group in combination with time-reversal ($\mathcal T$) symmetry, are of fundamental importance for TSMs and provide an important underpinning for the theory of TSMs. 

Over the past decade, the symmetry analysis of electronic energy bands has been supplemented with an equally powerful and consequential concept: band topology. Electronic phases can be sharply distinguished based on the topological properties of the ground state wavefunction, or, more precisely, the topology of the mapping from the Brillouin zone to the space of wave functions. 
In particular, insofar as TSMs are concerned, band crossings can typically be associated with a formal topological index, which is sensitive to the symmetries of the system, and this index gives rise to the notion of topological stability.

The implications of symmetry and topology are exemplified by two canonical examples of TSMs: the Weyl and Dirac semimetal. Even without any symmetry, a degeneracy of two energy bands may generically be expected to occur (\ie, without fine-tuning) at a point in the three-dimensional Brillouin zone~\cite{Herring37p365}. The dispersion away from the touching point is generically linear and this defines a Weyl point or Weyl node, which, as discussed in more detail in Section \ref{sec:weyl}, is characterized by a topological invariant called the Chern number~\cite{Wan11p205101}.  The Chern number protects the Weyl point in the sense that the degeneracy cannot be removed unless two Weyl points with opposite Chern number are brought to coincidence at the same momentum, by tuning Hamiltonian parameters or breaking translation symmetry, allowing them to hybridize. When $\mathcal T$ symmetry and inversion ($\mathcal P$) symmetry are present all energy bands are twofold degenerate and Weyl point degeneracies must come in pairs of opposite Chern number at the same momentum. Without further (crystalline) symmetries, however, the Weyl points are still allowed to hybridize, leading to an avoided crossing. To protect a fourfold-degenerate merger of two Weyl points, and thereby stabilize a Dirac semimetal phase, additional symmetries must present, such as rotation symmetry or nonsymmorphic space group symmetries. The stability of a rotation-symmetry protected Dirac point can be formally expressed in terms of a topological number built from energy band rotation eigenvalues~\cite{Yang15p165120}.

These considerations serve to illustrate the significance of both symmetry and topology for TSMs. In general, the quantized topological numbers associated with the stability of TSMs depend on the type of band crossing and symmetry class. In many cases the topological properties of TSMs can be diagnosed by calculating momentum space Berry phases and Wilson loops~\cite{Yu12p075119}, which have become standard tools of topological band theory. In other cases a formal topological number can be defined in terms of energy band symmetry quantum numbers. Many of these tools have been implemented in {\it ab initio} band structure calculations~\cite{ Bansil16p021004,Yu17p127202,Hirayama18p041002}. Density functional theory (DFT) within the single particle approximation provides a good description of weakly correlated electronic systems and has proven to be instrumental in finding material realizations of  TSMs. Recent illustrative  examples include the Weyl semimetal TaAs~\cite{Weng15p011029,Huang15p7373}, the ``triple-point'' semimetals MoP~\cite{Lv17p627} and WC~\cite{Ma18p349}, and the Dirac nodal line semimetal ZrSiS~\cite{Neupane16p201104}. The implementation of computational schemes to calculate topological quantities such as Berry phases has been particularly important for predicting TSMs, as they can be used to detect band crossings at generic non-high-symmetry momenta. 

We begin by reviewing the basic aspects of Dirac and Weyl semimetals in the next two sections. We then turn to a discussion of nodal line semimetals, after which we proceed to a survey of a number of more recently introduced families of TSMs. In the final section, before a summary and outlook, we consider some recent developments in the systematic search for new topological materials, in particular novel TSMs. In an appendix we include a table collecting candidate materials which have been proposed; some but not all are discussed in the text. 

Before we proceed, we feel compelled to point out the limitations inherent in reviewing a branch of condensed matter materials science as vast and diverse as TSMs. Our review covers a number of the recent developments but not all. In addition, the perspective and emphasis we choose is far from the only possible perspective one may adopt. In this regard, we note that a number of excellent longer and shorter reviews already exist in the general area of gapless topological materials~\cite{Turner13p293, Wehling14p1, Vafek14p83, Hasan15p014001, Weng16p303001, Fang16p117106, Yang16p1640003, Bansil16p021004, Jia16p1140, Yan17p337, Hasan17p289, Armitage18p015001, Yu17p127202, Yang18p1414631, Burkov18p359, Schoop18p3155, Bernevig18p041001, Hirayama18p041002}.

\section{DIRAC SEMIMETALS  \label{sec:dirac}}

The Dirac Hamiltonian plays a central role in the theory of topological materials. Its significance follows from the fact that in the presence of $\mathcal T$ and $\mathcal P$ symmetry a three-dimensional Dirac fermion appears at the topological phase boundary between a normal insulator and a topological insulator~\cite{Murakami07p356}. To illustrate this in more detail, consider the Dirac Hamiltonian in three dimensions given by
\begin{equation}
H({\bm k}) = \begin{pmatrix} m &  v {\bm k} \cdot \bm{\sigma} \\ v  {\bm k} \cdot \bm{\sigma} &-m  \end{pmatrix} = v  {\bm k} \cdot \bm{\sigma}  \tau_x + m \tau_z ,  \label{eq:dirac}
\end{equation}
where $\bm k =(k_x,k_y,k_z)$ is momentum, $v$ is a velocity (in condensed matter referred to as the Fermi velocity $v_{\mathrm F}$), and $m$ is a mass; $\bm{\sigma}=(\sigma_x,\sigma_y,\sigma_z)$ and $\bm{\tau}=(\tau_x,\tau_y,\tau_z)$ are the two sets of Pauli matrices, which can be viewed as spin and orbital degrees of freedom, respectively. The mass parameter $m$ controls the topological transition: when $m$ is tuned such that it changes sign, the topology of the ground state changes. A sign change of $m$ in Eq.~\eqref{eq:dirac} corresponds to a band inversion at a $\mathcal T$-invariant momentum and thus describes the transition between a normal insulator and a topological insulator. At the critical point of the transition, defined by $m=0$, the spectrum of \eqref{eq:dirac} is gapless at $\bk=0$, giving rise to a fourfold-degenerate point node with linear dispersion. As a result, at the critical point the low-energy electronic excitations are governed by a massless Dirac equation.  

This is reminiscent of the low-energy electronic structure of graphene, which is described by a Dirac equation in two dimensions~\cite{Neto09p109}. The experimental realization of graphene~\cite{Novoselov04p666} and the observation of the remarkable properties of Dirac electrons~\cite{Neto09p109} have made it a paradigmatic example of condensed matter realizations of Dirac physics, and have inspired the attempt to generalize the realization of low-energy Dirac fermions to three dimensions. The critical point of a topological transition provides one possible realization, however, as is clear from Eq.~\eqref{eq:dirac}, this requires a fine-tuning of Hamiltonian parameters to achieve a bulk gap closure. One possible yet experimentally challenging route is to tune the chemical compositions in materials such as Bi$_{2-x}$In$_x$Se$_3$, Pb$_{1-x}$Sn$_x$Se, Pb$_{1-x}$Sn$_{x}$Te, Bi$_{1-x}$Sb$_{x}$, and Hg$_{1-x}$Cd$_{x}$Se, for which the critical point corresponds to specific values of $x$. First-principles calculations have predicted that this type of Dirac semimetal should occur in the materials ZrTe$_5$~\cite{Weng14p011002} and SrSn$_2$As$_2$~\cite{Gibson15p205128}, but its experimental verification has proven difficult~\cite{Manzoni16p237601,Li16p550,Zhang17p15512,Mutch18p07898} as a result of sensitivity to factors such as temperature, pressure, and composition homogeneity.

The intrinsic instability of the Dirac point at the topological phase boundary has led to the search for a novel type of Dirac point stabilized by crystalline symmetries. Semimetallic phases hosting stable Dirac points define true Dirac semimetals. To date, two mechanisms have been identified for the realization of stable Dirac points, giving rise to two types of Dirac semimetals. The first kind, referred to as nonsymmorphic Dirac semimetals, relies on the nonsymmorphic nature of the crystal space group, enforcing the occurrence of Dirac points at high-symmetry points of the Brillouin zone boundary~\cite{Young12p140405}. The second kind, referred to as topological Dirac semimetals, relies on a band inversion in the presence of an $n$-fold rotational symmetry axis for $n=3, 4$, and $6$, where the Dirac points appear in pairs off a high-symmetry momentum~\cite{Wang12p195320}. 

The Dirac semimetals in three dimensions have strong spin-orbit coupling and, importantly, the Dirac points are stable in the presence of spin-orbit coupling. This should be contrasted with the case of graphene, which strictly speaking is a (topological) insulator when spin-orbit coupling is accounted for~\cite{Kane05p226801, Young15p126803}. In the remainder of this section, we discuss the three-dimensional Dirac semimetals in detail. 

\subsection{Symmetry-enforced and band inversion induced Dirac semimetals  \label{ssec:sym-invert}}

\begin{figure}[t]
\includegraphics[width=0.8\textwidth]{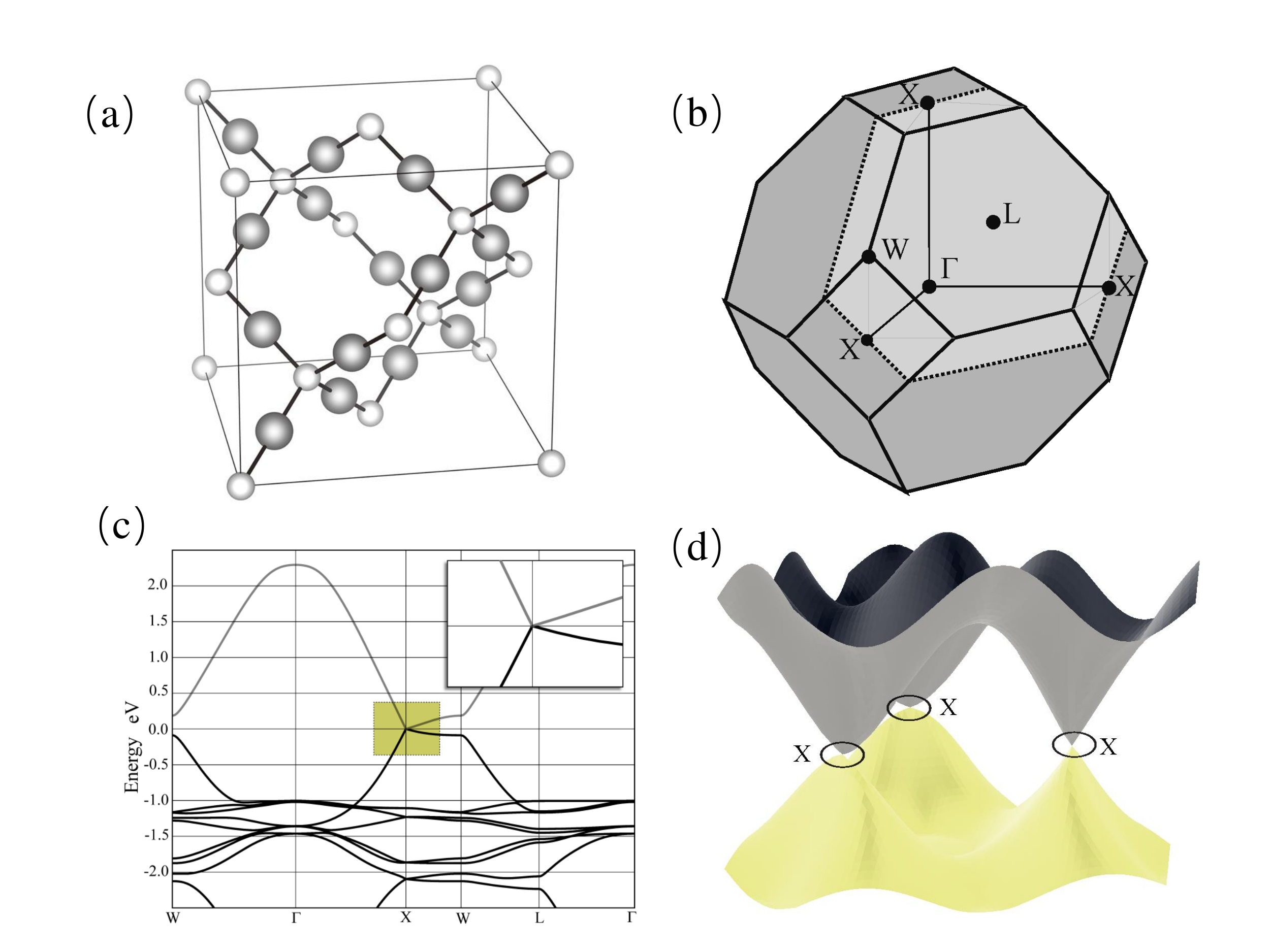}
\caption{ 
{\bf Symmetry-enforced Dirac semimetal.}  
A nonsymmorphic symmetry-enforced Dirac semimetal phase was theoretically proposed to exist in $\beta$-cristobalite ${\rm BiO_2}$, of which the SiO$_2$-type crystal structure is shown in (a). Here, bismuth atoms (light gray) are arranged on a diamond lattice, with oxygen atoms (dark gray) sitting midway between pairs of bismuth. Panel (b) shows the Brillouin zone of the FCC lattice; the plane highlighted in gray joins the three symmetry related $X$ points. Other high-symmetry points are also indicated. The band structure of BiO$_2$ in the $\beta$-cristobalite structure is shown in (c) and (d) shows the conduction and valence bands of $\beta$-cristobalite BiO$_2$ plotted as functions of momentum in the plane, which touch at the Dirac points. {\it Adapted from Ref.~\cite{Young12p140405}}
}
\label{fig:Dirac-A}
\end{figure}

The symmetry-enforced {\em nonsymmorphic} Dirac semimetal was first proposed by Young {\em et al.}~\cite{Young12p140405}, who examined the conditions under which the symmetries of the crystal space group mandate fourfold degeneracies with linear dispersion. It was found that such fourfold degeneracies, which are associated with four-dimensional irreducible (co)representations of the little group, can only occur in the case of nonsymmorphic space groups, and must be located at high-symmetry momenta on the Brillouin zone boundary. Based on this general analysis a full list of space groups which admit such four-dimensional irreducible representations at high-symmetry points was obtained~\cite{Zaheer14thesis}, revealing that 69 nonsymmorphic space groups can host symmetry-enforced Dirac points at a zone corner. 

To realize the nonsymmorphic Dirac semimetals in real materials, it is desirable that the fourfold-degenerate Dirac point is at or close to the Fermi level. Furthermore, the low-energy electrons should come only from the Dirac nodes, \ie there should be no other Fermi surfaces from other overlapping bands. Young {\etal} proposed $\beta$-cristobalite ${\rm BiO_2}$ as a symmetry-enforced Dirac semimetal~\cite{Young12p140405} that satisfies these criteria. The crystal structure of $\beta$-cristobalite ${\rm BiO_2}$ in space group $\bf 277$ ($Fd3m$) is shown in Fig.~\ref{fig:Dirac-A}(a). It is isostructural to $\beta$-cristobalite ${\rm SiO_2}$ which consists of silicon atoms on a diamond lattice with oxygen atoms placed midway along each silicon-silicon bond. The first-principles band structure of $\beta$-cristobalite ${\rm BiO_2}$ in Fig.~\ref{fig:Dirac-A}(c) clearly show fourfold-degenerate band crossings with linear dispersion at the $X$ points. The experimental confirmation of the Dirac semimetal phase in BiO$_2$ has remained illusive, however, due to its chemical instability. Subsequently, further materials candidates were identified in the distorted spinels, such as BiZnSiO$_4$, BiCaSiO$_4$, BiAlInO$_4$, and BiMgSiO$_4$~\cite{Steinberg14p036403}, a family of cluster compounds $A$Mo$_3X_3$ with $A=(\rm Na, \rm K,\rm Rb,\rm In,\rm Tl)$, and $X=(\rm Se, \rm Te)$ and HfI$_3$~\cite{Gibson15p205128}. Despite these predictions, this type of nonsymmorphic Dirac semimetals has not yet been experimentally realized. 

\begin{figure}[t]
\includegraphics[width=1\textwidth]{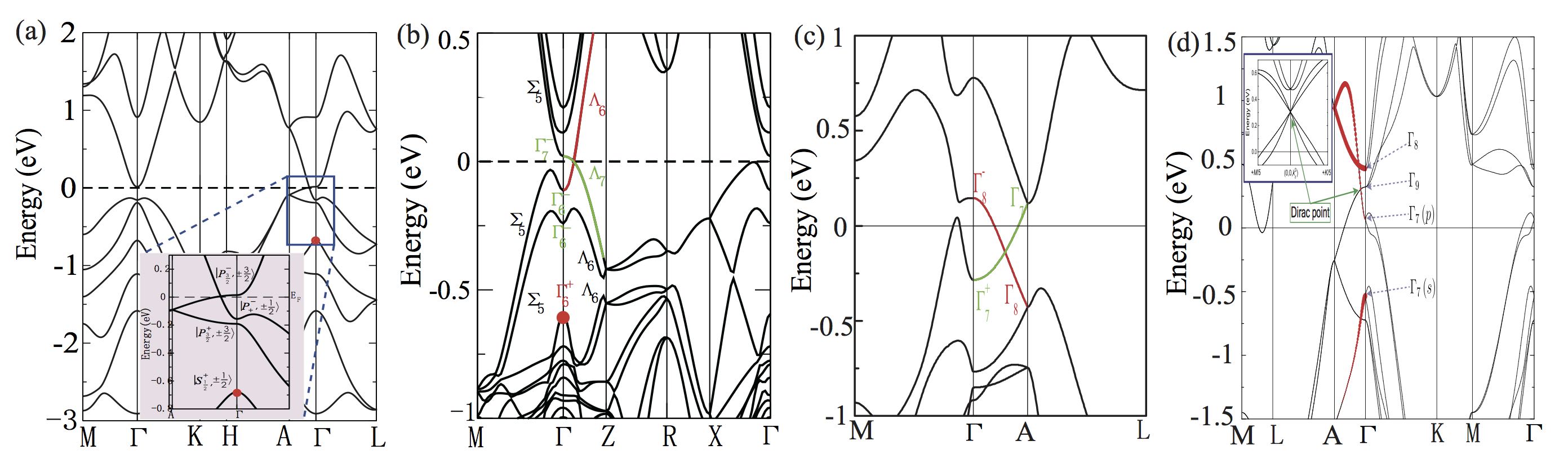}
\caption{ { \bf Band inversion induced Dirac semimetal. }
Shown are band structures of band inversion induced Dirac semimetals obtained from first-principles calculations. The Dirac points are located on the rotation axis and protected by rotation symmetry. Panels (a) and (b) show the experimentally confirmed Dirac semimetal materials Na$_3$Bi and Cd$_3$As$_2$ with space group $\bf 194$ ($P6_{3}/mmc$) and $\bf 137$ ($P4_{2}/nmc$), respectively. Panels (c) and (d) show the proposed hexagonal $ABC$ materials BaAgBi and LiZnBi with nonpolar and polar space groups $\bf 194$ ($P6_{3}/mmc$) and $\bf 186$ ($P6_{3}mc$), respectively.
{\it Adapted from Ref.~\cite{Wang12p195320,Wang13p125427,Du15p14423,Cao17p115203}.}
}
\label{fig:Dirac-B}
\end{figure}

The other type of three-dimensional Dirac semimetals, referred to as {\em topological} Dirac semimetals, arise due to symmetry-protected band crossings induced by a band inversion. While a band inversion generically results in avoided crossings, in the presence of rotation symmetry gap closures can occur when bands with different rotation eigenvalues cross on the rotation axis. On the rotational axis, in the presence of $\mathcal T$ and $\mathcal P$ symmetry, all twofold-degenerate bands can be labeled by representations of rotation axis little group. A crossing of two of these bands with different symmetry representations then leads to stable fourfold-degenerate Dirac points on the rotation axis. Yang and Nagaosa have systematically considered such accidental band crossings and classified the rotation symmetry-protected Dirac semimetals, finding that $n$-fold rotations $C_{n}$ with $n=3,4,6$ can stabilize band inversion induced Dirac points~\cite{Yang14p4898}. The stability of a such a Dirac point can be formally expressed in terms of a topological number constructed from energy band rotation eigenvalues~\cite{Yang15p165120}. Note that here the stability is of a weaker kind than the aforementioned symmetry-enforced Dirac semimetals, as the band-inversion can be removed by tuning parameters without breaking symmetries. 

The first material candidates were predicted by Wang {\etal}, who proposed a family of alkali pnictides A$_3$Bi (A=Na, K, Rb)~\cite{Wang12p195320} and Cd$_3$As$_2$~\cite{Wang13p125427} as the realizations of {\em topological} Dirac semimetals. The band structures of Na$_3$Bi and Cd$_3$As$_2$ in the presence of spin-orbit coupling are shown in Fig.~\ref{fig:Dirac-B}(a) and (b), exhibiting linear crossing on the rotation $z$ axis. 
Although Na$_3$Bi and Cd$_3$As$_2$ have different space group and composition, the low-energy electronic structure near $\Gamma$ giving rise to the Dirac points can be understood from the same effective Hamiltonian given by~\cite{Wang12p195320,Wang13p125427}
\begin{equation}
H(\bm k)=\varepsilon_0(\bm k) + \begin{pmatrix} M({\bm k}) & Ak_+ &0 & B^*({\bm k}) \\ Ak_- & -M({\bm k})  &B^*({\bm k}) &0 \\ 0&B({\bm k}) & M({\bm k}) & -Ak_- \\  B({\bm k}) & 0&-Ak_+ &-M({\bm k}) \end{pmatrix} , \label{eq:dirac-invert}
\end{equation}
where $k_\pm=k_x\pm ik_y$ and the functions $M(\bm k)$ and $B(\bm k)$ are defined as $M(\bm k)=M_0-M_1k_z^2-M_2(k_x^2+k_y^2)$ and $B(\bm k)=B k_z k^2_+$. The parameters $A$, $B$, and $M_{0,1,2}$ are material dependent. The band inversion and thus the presence of Dirac points are controlled by $M(\bm k)$. In particular, when the parameters $M_{0,1,2}$ are all positive (or all negative) the bands are inverted and a pair of Dirac points is located at momenta $k_z$ given by the solutions to the equation $M(k_z,0,0) = 0$, giving $k_z = \pm \sqrt{M_0/M_1}$. Expanding the Hamiltonian \eqref{eq:dirac-invert} around these points and keeping only the lowest order linear terms, the Dirac Hamiltonian can be readily obtained. Following the first-principles predictions of the Dirac semimetals in Na$_3$Bi and Cd$_3$As$_2$, angle-resolved photoemission spectroscopy experiments have confirmed the existence of Dirac points in these two materials~\cite{Liu14p864,Liu14p677}, making them the primary object of experimental studies addressing the special electronic and transport properties of three-dimensional Dirac semimetals.

In addition to these materials, a considerable group of materials with the ZrBeSi structure type has been proposed as topological Dirac semimetals~\cite{Gibson15p205128,Du15p14423}. Hexagonal ZrBeSi-type compounds have the space group as Na$_3$Bi and consist of layers of BeSi honeycomb nets separated by the Zr cations. An example is BaAgBi, for which the calculated band structure is shown in Fig.~\ref{fig:Dirac-B}(c). The physics of the band inversion giving rise to the Dirac points is similar to Na$_3$Bi. While all these materials are centrosymmetric, the topological Dirac semimetals can also exist in noncentrosymmetric crystals without inversion symmetry. A general analysis of such semimetals was presented by Gao {\etal}~\cite{Gao16p205109}, who showed that rotation axes with little groups isomorphic to the $C_{4v}$ and $C_{6v}$ point groups allow for stable fourfold-degenerate crossings. Material realizations of noncentrosymmetric Dirac semimetals were proposed in a family of hexagonal $ABC$ materials with LiGaGe-type structure and polar space group $\bf 186$ ($P6_{3}mc$). The $B$ and $C$ atoms occupy the sites of a Wurtzite structure and $A$ forms the interstitial site. Examples include LiZnBi~\cite{Cao17p115203}, shown in Fig.~\ref{fig:Dirac-B}(d), CaAgBi~\cite{Chen17p044201}, and SrHgPb~\cite{Gao18p106404}. Another noncentrosymmetric Dirac semimetal is Cd$_2$As$_3$ in space group $\bf 110$ ($I4_{1}cd$) which has a different vacancy order than the centrosymmetric $P4_{2}/nmc$ phase. 

The Dirac equation given by Eq.~\eqref{eq:dirac} describes excitations with linear and isotropic dispersion. In the case of free elementary particles, for which the Dirac equation was introduced, this form follows from the requirement of Lorentz invariance. In contrast, the Dirac theory in crystals is not constrained by Lorentz invariance, allowing for a more general form of the Dirac Hamiltonian. This is clear from Eq.~\eqref{eq:dirac-invert}, for instance, which implies anisotropic Dirac points. More importantly, however, the general classification of Dirac semimetal phases obtained by Yang and Nagaosa, which is based on the dispersion at the Dirac points, demonstrated that the dispersion can be quadratic or cubic, depending on the symmetry quantum numbers of the bands which cross to form the Dirac point~\cite{Yang14p4898}.  Liu and Zunger proposed a class of quasi-one-dimensional transition-metal monochalcogenides compounds $\rm{A^{I}(MoX^{VI})_{3}}$ ($\rm{A^{I}}$=Na, K, Rb, In, Tl; $\rm{X^{VI}}$ = S, Se, Te) as candidates of an ideal cubic Dirac semimetals~\cite{Liu17p021019}. 

\subsection{Antiferromagnetic Dirac semimetals \label{ssec:dirac-AFM}}

The Dirac semimetals considered in the previous subsection are all nonmagnetic preserving $\mathcal T$ symmetry. Recent progress has been achieved in extending the analysis including magnetic systems. The possibility of stable Dirac points in magnetic systems was addressed by Tang {\etal}~\cite{Tang16p1100}, showing that an antiferromagnetic ordering of spins that preserves $\mathcal{PT}$ can protect Dirac point. With the combined symmetry, the bands becomes twofold-degenerate at each momentum and a crossing of bands, protected by a crystal symmetry, can give rise to a fourfold-degenerate Dirac point. First-principles calculations predicted magnetic CuMnAs as a antiferromagnetic Dirac semimetal~\cite{Tang16p1100}.

Subsequent work examined the generalization of symmetry-enforced Dirac semimetals to (antiferro-)magnetic systems, both in two~\cite{Young17p186401} and three dimensions~\cite{Wang17p00896}. In these systems the existence of Dirac points is mandated by the symmetries of the magnetic space group and the symmetry-enforced degeneracies occur at high-symmetry points of the (magnetic) Brillouin zone. Here, an important role is played by an anti-unitary symmetry composed of time-reversal $\mathcal T$ and a half-translation, \ie a translation conjugate to the magnetic order vector. A systematic symmetry-based analysis of band topology in all 1651 magnetic space groups was performed by Watanabe~{\etal}, who identified magnetic space groups which, for specific electron filling constraints, necessarily imply a symmetry protected gapless state~\cite{Watanabe18p8685}. To date, further materials have been identified, including EuCd$_2$As$_2$~\cite{Hua18p02806} protected by $C_{3}$ rotation symmetry and CeSbTe proposed to host Dirac and Weyl states~\cite{Schoop18peaar2317}.

\section{WEYL SEMIMETALS  \label{sec:weyl}}

For a basic understanding of the Weyl semimetal it is helpful to return to the Dirac Hamiltonian of Eq.~\eqref{eq:dirac}. In the massless case, when $m=0$, the Dirac Hamiltonian decouples into two separate equations given by $\pm v \bm k\cdot \bm \sigma$, where each equation describes a two-component chiral Weyl fermion with chirality $\pm 1$. In general, a two-component Weyl fermion described by the Weyl equation can arise in solids when two non-degenerate energy bands touch at a point $\bk_0$ in momentum space. Clearly, this cannot occur when Kramers degeneracy holds at every momentum $\bm k$, and therefore $\mathcal T$ and $\mathcal P$ symmetry cannot both be present; at least one of these two symmetries must be broken. In the absence of any crystalline symmetries a linear touching of non-degenerate bands can occur at a generic momentum $\bk_0$ in the three-dimensional Brillouin zone. One may expand the Hamiltonian about the degeneracy point to linear order in momentum $\delta \bm k = \bm k - \bm k_0$ to obtain
\begin{equation}
 H(\delta \bm k)=\varepsilon_{0}(\bm k_0)  + \bm v_0  \cdot   \delta \bm k+ \sum_{i,j=x,y,z} v_{ij} \delta k_i \sigma_j ,   \label{eq:weyl-expand} 
\end{equation}
where $\sigma_{x,y,z}$ are again the Pauli matrices and $\bm v_0, \bm v_{x,y,z} $ characterize the band dispersion near the touching point. The significance of the second terms in \eqref{eq:weyl-expand} will be discussed in Section~\ref{ssec:Weyl-C=2,3}, here we focus attention on the third term, which encodes the topological structure of the Weyl point. 

To see this, let us first note that when $v_{ij} = v\delta_{ij}$ one recovers $v \delta \bm k\cdot \bm \sigma$, indicating that \eqref{eq:weyl-expand} indeed describes a Weyl fermion. The topology of Weyl fermions follows from the fact that Weyl points are monopoles of momentum space Berry curvature. The Berry curvature $F^{ab}_n(\bm k) $ of a particular energy band labeled by $n$ is the field strength of the Berry connection $\mathcal A^a_n(\bm k)$ and is defined as 
\begin{equation}
\mathcal F^{ab}_n(\bm k) =\nabla^{a} \mathcal A^b_n(\bm k) -\nabla^{b} \mathcal A^a_n(\bm k),  \quad \mathcal A^a_n(\bm k) = -i \langle \psi_n(\bm k)  | \nabla^{a} | \psi_n(\bm k) \rangle , \quad 
\end{equation}
where $\nabla^{a} \equiv \partial/\partial k_a$ and $a,b=x,y,z$. Defining the momentum space ``magnetic field'' as $\mathcal B^{a}_n(\bm k)  = \epsilon_{abc}\mathcal F^{bc}_n(\bm k)/2$ the topological nature of a Weyl point at momentum $ \bm k_0$ can be expressed as the quantized monopole density $\frac{1}{2\pi}\bm \nabla \cdot \boldsymbol{\mathcal B}_n(\bm k) = q \delta(\bm k- \bm k_0)$. For the general Weyl Hamiltonian given by \eqref{eq:weyl-expand} one finds $q = \text{sgn}(\text{Det} \, v_{ij})$. In the particular case $H(\delta \bm k) =\pm v \delta \bm k\cdot \bm \sigma $ this reduces to $q = \pm 1$ and the momentum space magnetic field of the valence band is given by $ \boldsymbol{\mathcal B}(\delta\bm k) =\pm  \delta \bm k /2| \delta \bm k|^3$. Integrating $ \boldsymbol{\mathcal B}(\delta \bm k) $ over a Fermi surface sheet surrounding the Weyl point defines the Chern number $C = \int  \boldsymbol{\mathcal B} \cdot d\bm \Omega /2\pi$ and equals the monopole charge $C=\pm 1$. (In general $C=q$.)

Two important consequences directly follow from this. First, since the Nielsen-Ninomiya theorem \cite{Nielsen83p389} states that the total monopole charge integrated over the Brillouin zone must vanish, the sum over all chiral fermions must be zero (see also Section~\ref{ssec:spin-j}). In a system with only chiral Weyl fermions the Weyl points must come in pairs of opposite monopole charge. (This is, for instance, why the massless Dirac equation of Eq.~\eqref{eq:dirac} decouples into two Weyl equations with opposite chirality.) Second, since $\mathcal T$ symmetry implies $\mathcal F^{ab}_n(\bm k) = - \mathcal F^{ab}_n(-\bm k) $ and $\mathcal P$ symmetry implies $\mathcal F^{ab}_n(\bm k) =  \mathcal F^{ab}_n(-\bm k) $, the Berry curvature vanishes when both symmetries are present. Hence, the symmetry properties of the Berry curvature provide a more formal argument way of demonstrating that the existence of Weyl points requires at least one of these symmetries to be broken. 

The most important implication of these considerations is the topological stability of Weyl points: since monopoles of Berry curvature cannot be removed unless they are brought to coincidence with monopoles of opposite charge, Weyl points are topologically stable as long as they remain separated in momentum space. In particular, a topological phase transition from a  Weyl semimetal phase to an insulating phase requires tuning Hamiltonian parameters to bring pairs of Weyl points together, allowing them to hybridize and annihilate; the degeneracy of a single Weyl point cannot be removed. A more heuristic (and perhaps more intuitive) way to state this result is to say that for the Weyl Hamiltonian $H(\delta \bm k) = v \delta \bm k\cdot \bm \sigma $, or its more general version $H(\delta \bm k) =\sum_{ij}  v_{ij} \delta k_i \cdot  \sigma_j $, it is not possible to write down a mass term, since all Pauli matrices have been already been used. 


\begin{figure}[t]
\includegraphics[width=\textwidth]{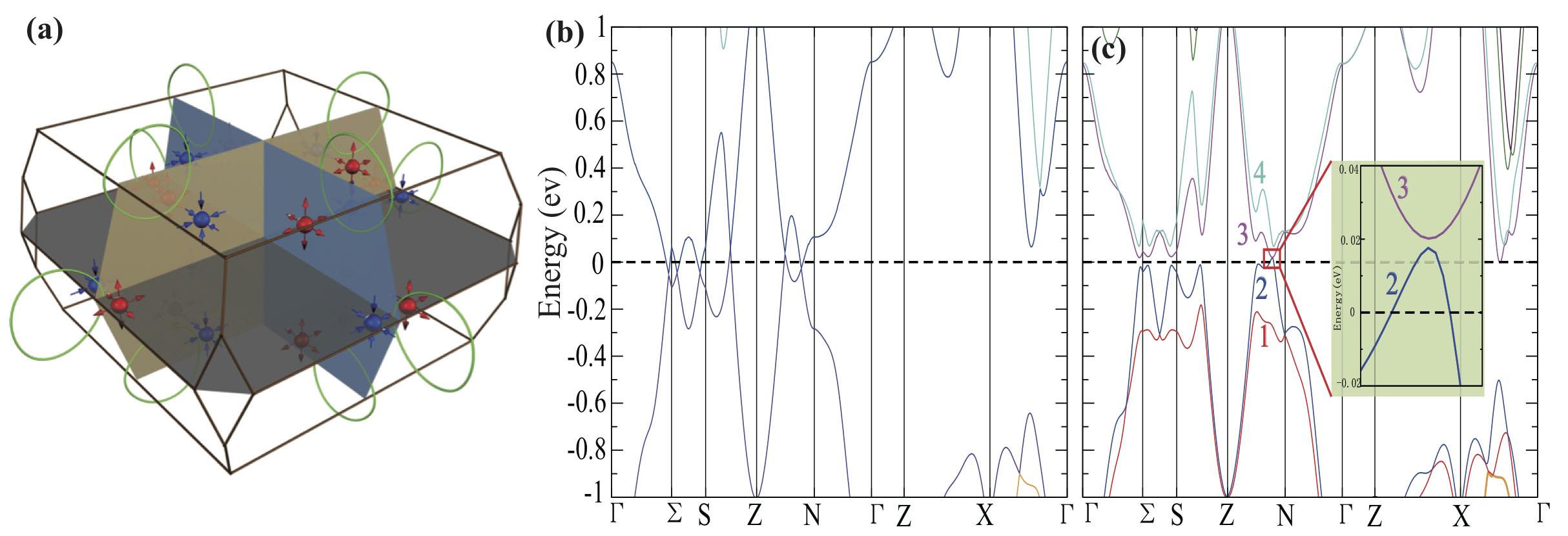}
\caption{ { \bf Weyl semimetal. }  
Panel (a) shows a three-dimensional view of the Weyl points in TaAs, indicated by red and blue dots. The colors and arrows represent the monopole charge of the Weyl points. The green rings represent the nodal rings which are present in the band structure when spin-orbit coupling is not taken into account. Panels (b) and (c) show the calculated band structures of TaAs without and with including the spin-orbit coupling, respectively. Note that the Weyl points of TaAs exist off the high-symmetry lines and are not apparent from (c).
{\it Adapted from Ref. \cite{Weng15p011029}.}}
\label{fig:weyl}
\end{figure}

\subsection{Realizations of $\mathcal{T}$-broken and $\mathcal P$-broken Weyl semimetals \label{ssec:Weyl-T,I}}

As discussed above, the Weyl semimetal phase relies on breaking either $\mathcal T$ or $\mathcal P$ symmetry (or possibly both). This has inspired two broad categories of proposals for realizing Weyl semimetals. The first category comprises magnetic candidate materials, which includes the first proposal for a Weyl semimetal by Wan {\etal}~\cite{Wan11p205101}. Wan {\etal} considered a class of magnetic pyrochlores $A_2$Ir$_2$O$_7$, where $A$ is Y or rare earth element Eu, Nd, Sm, or Pr, and predicted based on first-principles calculations that an "all-in, all-out" noncollinear magnetic order gives rise to Weyl points. In addition to the magnetic pyrochlores, the ferromagnetic half-metal HgCr$_2$Se$_4$ was predicted to host Weyl points~\cite{Xu11p186806}. More recently, a class of magnetic Co-based magnetic Heusler compounds has been proposed as Weyl semimmetals~\cite{Wang16p236401, Chang16p38839}, in addition to the antiferromagnetic Mn$_3$Sn family of materials~\cite{Yang17p015008} and layered half-metal Co$_3$Sn$_2$S$_2$~\cite{Xu18p235416}. These magnetic Weyl semimetals exhibit giant anomalous Hall effect in experiment~\cite{Nayak16pe1501870,Kiyohara16p064009,Liu18p1,Wang18p3681}, which provides strong evidence for the existence of Weyl points. Another route to the realization of $\mathcal T$-breaking Weyl semimetal is based on splitting the two Weyl nodes which form Dirac point, see Eq.~\eqref{eq:dirac}. To achieve this, Burkov and Balents proposed a heterostructure of alternating layers of magnetically doped topological insulator and normal insulator~\cite{Burkov11p127205}. A similar proposal is based on an CdO/EuO superlattice, in which case ferromagnetically ordered EuO is the normal insulator~\cite{Zhang14p096804}. 

The second category of proposals for Weyl semimetal phases is focused on noncentrosymmetric systems with strong spin-orbit coupling. Compared to magnetic materials, nonmagnetic Weyl semimetals have the advantage that the electronic structure is more easily probed by angle-resolved photoemission spectroscopy. Inversion symmetry breaking Weyl semimetals differ from $\mathcal T$ breaking Weyl semimetals by the minimum number of Weyl points present in the band structure. This can be understood from the effect of $\mathcal T$ symmetry: a Weyl point at $\bm k_0$ must have a partner at $-\bm k_0$ with the same chirality. Since the sum of chiralities of all Weyl points must be zero, this implies the existence of two more Weyl points and leads a minimal number of four Weyl points (assuming other types of chiral fermions are absent, see Section \ref{sec:generalized}). 

Different routes were explored to realize the $\mathcal T$-invariant Weyl semimetal phase. One route made use of Murakami's theory of the phase transition between the topological insulator and normal insulator in noncentrosymmetric materials, as such phase transition must necessarily occur via an intermediate Weyl semimetal phase~\cite{Murakami07p356}. An example of such proposal is the solid solution LaBi$_{1-x}$Sb$_x$Te$_3$ of the topological insulator LaBiTe$_3$ and normal insulator LaSbTe$_{3}$~\cite{Liu14p155316}. Another route, which is not unrelated to the former, has led to proposals based on the heterostructure principle, where layers of topological insulator and normal insulator are stacked to form a superlattice~\cite{Burkov11p235126, Zyuzin12p165110, Halasz12p035103}. There is, however, no experimental realization of the above proposals so far. 

A decisive advance came in 2015, when the observation of a $\mathcal T$-invariant Weyl semimetal phase was reported in TaAs and related materials~\cite{Xu15p748,Xu15p613,Lv15p031013,Lv15p724,Yang15p728,Bernevig15p698} based on earlier theoretical predictions~\cite{Weng15p011029, Huang15p7373}.  The band structure of TaAs is shown in Fig.~\ref{fig:weyl}. The experimental discovery of Weyl fermions in TaAs cemented the field of TSMs and fueled the subsequent quest for new Weyl and other semimetal phases.  

Ideal Weyl semimetals should have the following properties: {\it (i)} the Weyl points sit at the Fermi energy; {\it (ii)} there are no other band structure features, such electron or hole pockets, overlapping in energy; and {\it (iii)} the Weyl points have a relatively large separation in momentum space.  A Weyl semimetal satisfying criteria {\it (i)} and {\it (ii)} was proposed in the strained HgTe material class and in a CuTlTe$_2$ family of materials~\cite{Ruan16p226801}. A large separation of Weyl points is useful for experimental detection of the surface Fermi arc states and in this sense WP$_2$ and MoP$_2$~\cite{Autes16p066402}, as well as Ta$_3$S$_2$~\cite{Chang16p1600295} and CuTlTe$_2$ are promising candidates. For experimental studies it is furthermore desirable to have a system with a small number of Weyl points and with this in mind material candidates hosting the minimum number of four Weyl points were proposed, notably including the ternary compound TaIrTe$_4$~\cite{Koepernik16p201101} and MoTe$_2$~\cite{Wang16p056805}.

In the introductory discussion of this section we have mentioned that Weyl points may be annihilated by merging two partners with opposite chirality. This is enabled by the fact that a twofold band touching can occur at generic momenta in the Brillouin zone and is not necessarily pinned to one particular point. It is, however, possible for a (single) Weyl point to occur at a $\mathcal T$ invariant momentum. Kramers degeneracy then pins the Weyl fermion to the $\mathcal T$ invariant momentum, leading to an irremovable Weyl point. Such Weyl points are referred to as Kramers-Weyl fermions and can only occur in the 65 chiral space groups which lack inversion, mirror, and any other roto-inversion symmetries~\cite{Chang16p07925}. Several material candidates were proposed, including Ag$_2$Te and Ag$_3$BO$_3$~\cite{Chang16p07925}.

The breaking of $\mathcal P$ symmetry in nonmagnetic Weyl semimetals suggests that they may have interesting nonlinear optical properties, such as second harmonic generation and shift current, and these nonlinear optical responses have recently attracted much recent attention~\cite{Morimoto16p245121,deJuan17p15995,Wu17p350,Ma17p842,Yang17preprint,Flicker18preprint}. In particular, this attention has focused on the contribution of the Weyl fermions to the nonlinear optical response. Remarkably, the Weyl semimetal systems TaAs, TaP, and NbAs were found to have the largest the largest second harmonic generation ever recorded~\cite{Wu17p350}. 


\subsection{Type-II and multi-Weyl semimetals  \label{ssec:Weyl-C=2,3}}

\begin{figure}[t]
\includegraphics[width=1.0\textwidth]{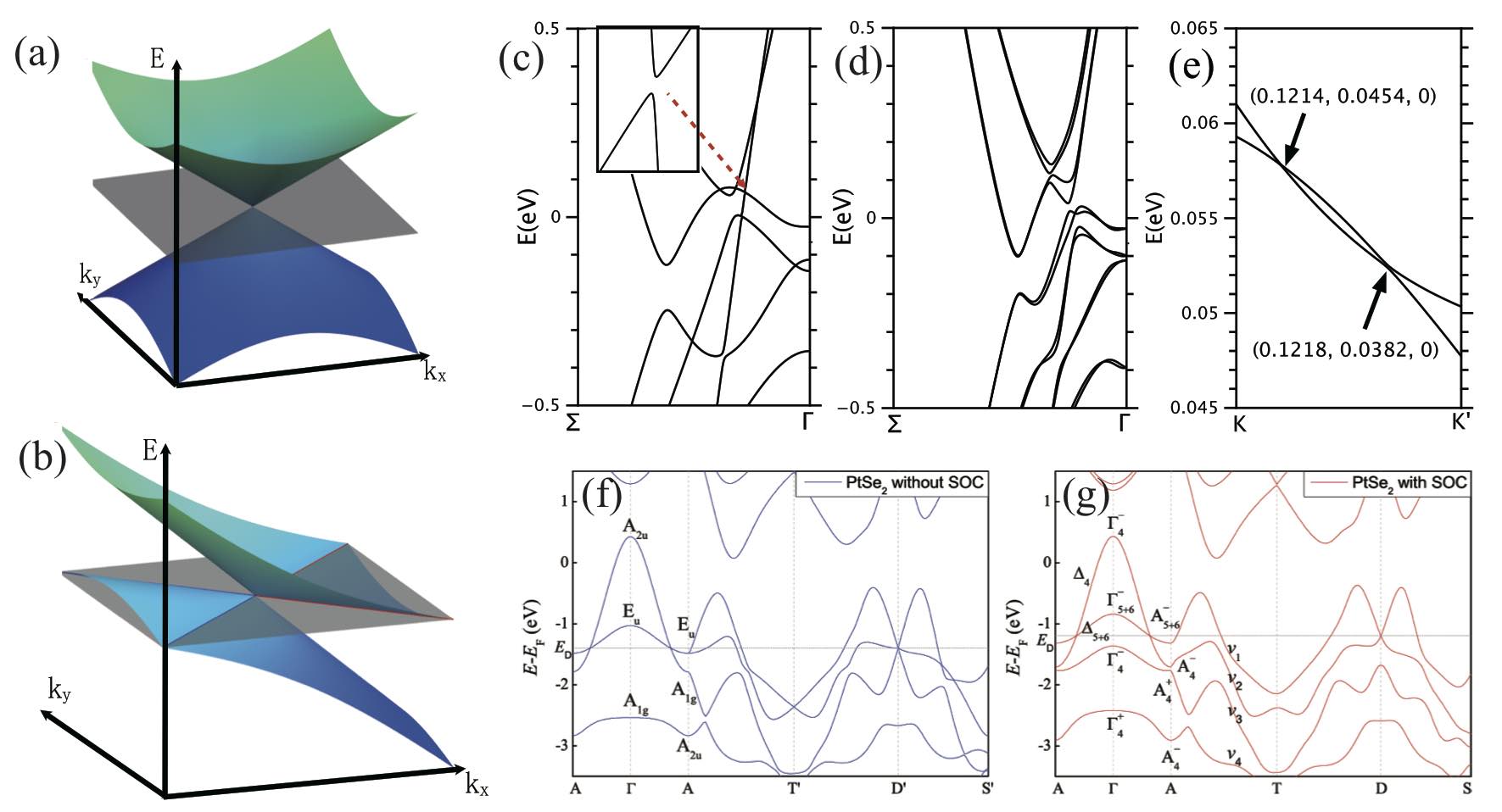}
\caption{{\bf Type-II topological semimetals.}  Panels (a) and (b) show the energy dispersion and Fermi surface of Type-I and Type-II Weyl/Dirac semimetals, respectively. Type-I Weyl/Dirac point with a point-like Fermi surface. A type-II Weyl/Dirac point appears as the contact point between electron and hole pockets. The grey plane corresponds to the position of the Fermi level, and the blue (red) lines mark the boundaries of the hole (electron) pockets. Panels (c) and (d) show the band structures of WTe$_2$ without and with SCO. Panel (e) shows one of the four pairs of Weyl points in WTe$_2$. Panels (f) and (g) show the band structures of PtSe$_2$ without and with spin-orbit coupling. {\it Adapted from Refs.\,\cite{Soluyanov15p495,Huang16p121117}}
}
\label{fig:weyl-type-II}
\end{figure}

Let us now return to Eq.~\eqref{eq:weyl-expand} describing the general form of the Hamiltonian near a Weyl point. We have shown that the third term encodes the topological property of the Weyl point (\ie the monopole charge) and in the simplest case takes the form an isotropic Weyl fermion. The second term, given by $\bm v_0  \cdot   \delta \bm k$, is important for the dispersion of the Weyl points and gives rise to a qualitative distinction between two types of Weyl points, called type-I and type-II Weyl points~\cite{Soluyanov15p495}. The phases which host these Weyl points are referred to as type-I and type-II Weyl semimetals. To understand this distinction, consider the energy spectrum corresponding to Eq.~\eqref{eq:weyl-expand} given by the two branches (ignoring the constant contribution)
\begin{equation}
\varepsilon_{\pm}(\delta \bm k)= \bm v_0  \cdot   \delta \bm k\pm \sqrt{\sum_{i,j=x,y,z}\delta k_i  (vv^{\mathrm T})_{ij}\delta k_j  }  \equiv T(\delta \bm k)\pm U(\delta \bm k) . \label{eq:energy-weyl} 
\end{equation}
Here, $T(\delta \bm k)$ and $U(\delta \bm k)$ may be viewed as the kinetic and potential part of the energy spectrum. Weyl points of type-II are then defined by the condition that for some direction $\delta \hat{ \bm k} = \delta \bm k/|\delta \bm k|$ in momentum space the kinetic energy exceeds the potential energy: $T(\delta \hat{ \bm k})> U(\delta \hat{ \bm k})$. This is associated with a qualitatively different structure of the spectrum, as it necessarily implies open electron and hole pockets when the Fermi energy is at the Weyl point. Instead, the Fermi surfaces of type-I Weyl points are closed and consist of a single point when the Fermi level is at the band crossing point. This is illustrated in Figs.~\ref{fig:weyl-type-II} (a) and (b), which show a type-I and type-II Weyl point, respectively. An intuitive way to picture this difference is to view a type-II Weyl point as a Weyl fermion with an over-tilted cone. Clearly, the tilting of the cone is due to the vector $\bm v_0$ in Eq.~\eqref{eq:weyl-expand} and can thus be called a tilt-vector.

The Weyl points observed in the TaAs material class are examples of type-I fermions. Soluyanov {\etal}, who pointed out the existence of these two distinct types of Weyl fermions, proposed WTe$_2$ as a possible material candidate to host Weyl points of type-II. The electronic band structure (in the presence of spin-orbit coupling) is reproduced in Fig.~\ref{fig:weyl-type-II} (d) and Fig.~\ref{fig:weyl-type-II} (e) shows the crossing of energy bands defining type-II Weyl points. Following this proposal, more material candidates were identified based on first-principles calculations, including MoTe$_2$~\cite{Wang16p056805}, the WP$_2$ material family~\cite{Autes16p066402}, TaIrTe$_4$~\cite{Koepernik16p201101}, Ta$_3$S$_2$~\cite{Chang16p1600295}, and LaAlGe~\cite{Xu16p07318}. 

Given the intimate relation between Weyl and Dirac fermions it is not surprising that the classification of Weyl fermions into type-I and type-II can also be applied to Dirac semimetals. From this perspective, the experimentally observed Dirac semimetals Na$_3$Bi and Cd$_3$As$_2$ mentioned in Section~\ref{ssec:sym-invert} realize type-I Dirac fermions, as may be checked from Eq.~\eqref{eq:dirac-invert}. The realization of a Type-II Dirac semimetal was predicted in the family of transition-metal icosagenides $MA_3$ with $M=(\rm V, \rm Nb, \rm Ta)$ and $A=(\rm Al, \rm Ga, \rm In)$~\cite{Chang17p026404} and in the PtTe$_2$ material family~\cite{Huang16p121117}. The band structure of PtTe$_2$ exhibiting type-II Dirac points is shown in Fig.~\ref{fig:weyl-type-II} (g).

The distinction between type-I and type-II Weyl fermions only arises in the context of condensed matter realizations of Weyl fermions. As alluded to in Section~\ref{sec:dirac}, this is due to the lack of Lorentz invariance in condensed matter systems and leads to an additional categorization of Weyl points based on their dispersion. As shown by Fang {\etal}~\cite{Fang12p266802}, apart from linear dispersion, Weyl fermions realized in crystals can have quadratic or cubic dispersion, depending on the rotation eigenvalues of the band which cross at the Weyl point. Notably, such ``multi-Weyl'' points have a monopole charge $q$ larger than that of the ordinary linear Weyl fermion. In particular, Weyl points with quadratic and cubic dispersion have monopole charge $q=\pm 2$ and $q= \pm 3$, respectively. The proposed ferromagnetic Weyl semimetal HgCr$_2$Se$_4$~\cite{Xu11p186806} is an example of a ``multi-Weyl'' semimetal, hosting Weyl points with quadratic dispersion. In addition, SrSi$_2$ was proposed as a charge $q=\pm 2$ Weyl semimetal~\cite{Huang16p1180}.


\section{NODAL LINE SEMIMETALS}

In this section, we review another class of topological semimetals referred to as topological nodal line semimetals. Nodal line semimetals feature gapless excitations along one-dimensional nodal lines formed from the linearly touching conduction and valence bands~\cite{Horava05p016405, Heikkia11p59, Burkov11p235126}. Nodal line degeneracies protected by crystalline symmetries can generically occur when bands with different crystal symmetry eigenvalues cross along a rotational axis or on a mirror- or glide-invariant plane of the BZ. In contrast to this type of nodal lines, nodal lines can occur due to band topology, in which case they are associated with a topological invariant. 

A variety of topological nodal line semimetal classes have been identified, which can be distinguished based on characteristics such as topological invariants~\cite{Fang15p081201}, degeneracy of the band crossing~\cite{Gao16p205109}, Fermi surface geometry~\cite{He18p053019, Kim18p1807.08523}, and the linking structure of multiple nodal lines~\cite{Bzduek16p75}. In this section, we review the characteristic features and material realizations of representative examples, which include $\mathbb{Z}_2$ Berry phase nodal line semimetals, $\mathbb{Z}_2$ monopole nodal line semimetals, and $\mathbb{Z}$ mirror- or glide-symmetric nodal lines. More comprehensive reviews of nodal line semimetals can be found in Refs.~\cite{Fang16p117106, Yu17p127202}.

\subsection{$\mathbb{Z}_2$ Berry phase nodal line semimetals}

This class of nodal line semimetals hosts fourfold-degenerate  Dirac nodal lines (including spin degree of freedom) in momentum space characterized by the $\mathbb{Z}_2$ quantized Berry phase. In the limit of vanishing spin-orbit coupling, assuming both time-reversal $\mathcal{T}$ and inversion $\mathcal{P}$ symmetry are present, the (doubly) Kramers degenerate bands can cross without opening a band gap along a one-dimensional line; this crossing is protected by a nontrivial $\mathbb{Z}_2$ quantized Berry phase, resulting in the formation of a fourfold-degenerate Dirac nodal line~\cite{Kim15p036806}. The $\mathbb{Z}_2$ Berry phase can be readily evaluated from the parity eigenvalues of all the occupied Bloch states at the time-reversal invariant momenta. The Berry phase  $\Omega(C_{abcd})$ corresponding to a closed loop $\mathcal{C}_{abcd}$ which visits four time-reversal invariant momenta (TRIMs) $\Gamma_{a,b,c,d}$ is represented in terms of the parity eigenvalues as $\Omega (\mathcal{C}_{abcd}) = \log(\xi_a \xi_b \xi_c \xi_d)/i$.  Here, $\xi_a = \prod_n \xi_n(\Gamma_a) = \pm 1$ are the parity eigenvalues of the occupied Bloch states at the eight time-reversal invariant momenta $\Gamma_a$, where $a = 1, \cdots, 8$. When $\Omega (\mathcal{C}_{abcd}) = \log(-1) /i = \pi$, an odd number of Dirac nodal lines must pierce any interior surface $\mathcal {S}_{abcd}$ that has the time-reversal invariant path $\mathcal{C}_{abcd}$ as its boundary. The $\mathbb{Z}_2$ Berry phases defined in terms of the parity eigenvalues, and corresponding to different surfaces $\mathcal {S}_{abcd}$, are quite similar to the (strong and weak) topological indices $(\nu_0;\nu_1\nu_2\nu_3)$ characterizing topological insulators in three dimensions~\cite{Fu07p106803, Fu07p45302}. We comment on this connection in more detail below. The $\mathbb{Z}_2$ invariant of Dirac nodal line semimetals given by the Berry phase formula can also be understood as the topological invariant characterizing one-parameter families of Hamiltonians in class AI~\cite{Teo10p115120}.

\begin{figure}[tb!]
\includegraphics[width=1.0\textwidth]{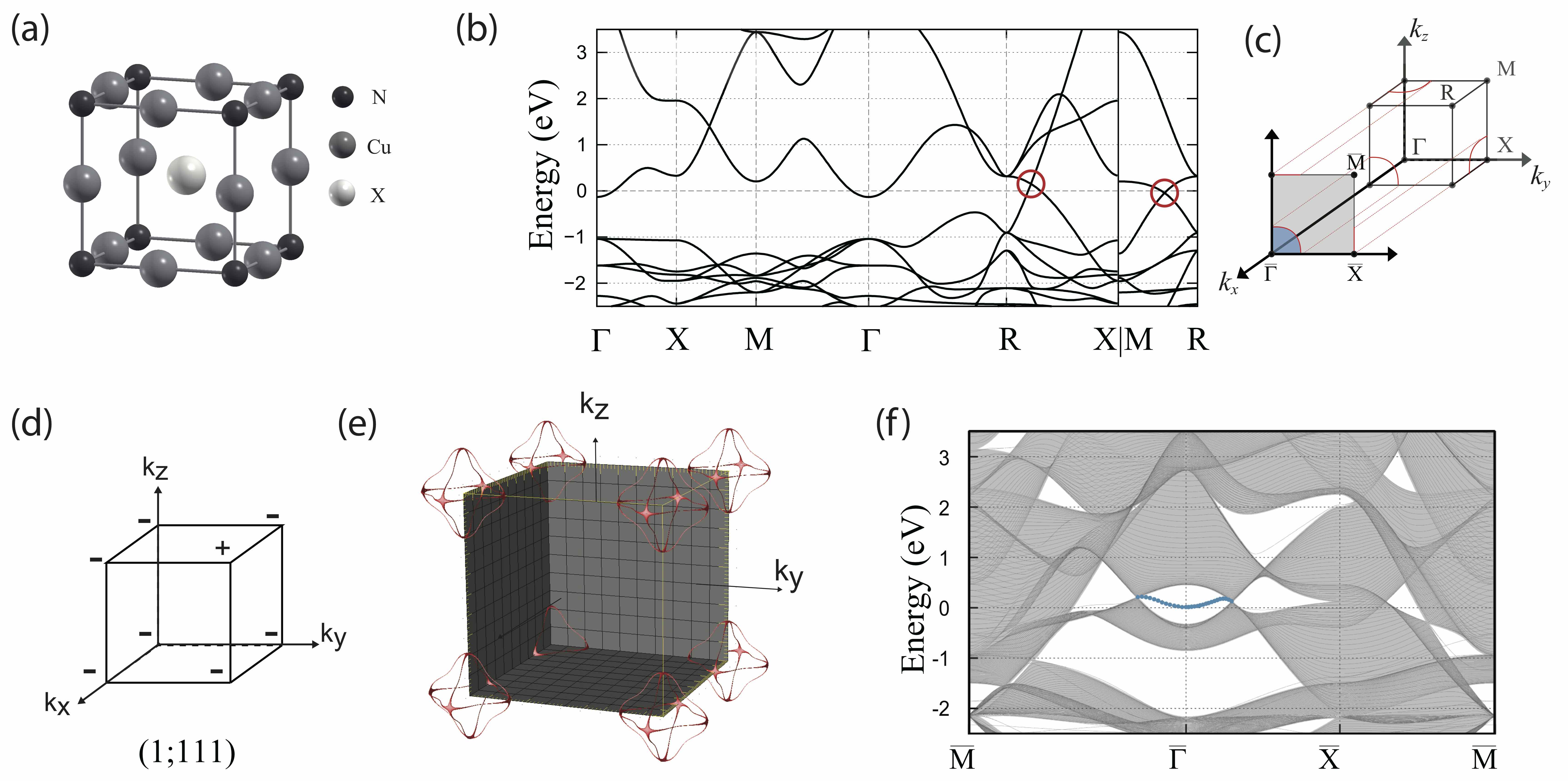}
\caption{ {\bf Nodal line semimetal in Cu$_3$N$X$.}
 (a) Anti-perovskite structure of Cu$_3$N$X$, where $X$ = Pd or Zn. (b) Band structure of Cu$_3$NPd. The crossing points indicated by red circles correspond to the points that comprise a Dirac nodal line.  (c) Bulk and surface BZs. (d) Parity eigenvalues evaluated at eight time-reversal invariant points of Cu$_3$NZn. (e) Nodal lines of Cu$_3$NPd in momentum space. (f) Surface energy spectrum of Cu$_3$NZn. 
 {\it Adapted from Ref.~\cite{Kim15p036806}}
}
\label{fig:yk_fig2}
\end{figure}

The representation of the $\mathbb{Z}_2$ Berry phase in terms of the band parity eigenvalues establishes an interesting link between the nodal line semimetals and the time-reversal invariant topological insulators with strong spin-orbit interaction. Indeed, in centrosymmetric topological insulators the $\mathbb{Z}_2$ topological indices can similarly be represented in terms of the parity eigenvalues $\xi_a$ \cite{Fu07p45302}. In both cases the topological indices constructed from the parity eigenvalues signal the presence of a band inversion. As a result, a $\mathbb{Z}_2$ Berry phase Dirac nodal line semimetal is turned into a $\mathbb{Z}_2$ topological insulator when spin-orbit coupling is activated, provided spin-orbit coupling does not undo the band inversion at a TRIM, thus changing the parity eigenvalues of the occupied Bloch states. 

Apart from establishing a conceptual connection to $\mathbb{Z}_2$ topological insulating phases, the representation of $\mathbb{Z}_2$ Berry phase in terms of parity eigenvalues also provides important insight into ways to realize nodal line semimetals in real materials.  When bands are inverted such that the parity eigenvalues change at a TRIM, the band inversion guarantees the presence of the nodal line. Furthermore, when the band inversion could be tuned via impurity doping or strain \cite{Kim15p036806, Zhao16p195104, Xu18p161111}, the size of the nodal line can be engineered, providing a potential route to realize strongly-correlated nearly-flat bands localized on the surface~\cite{Kim15p036806, Volovik15p014014, Heikkilae16p123}.   This insight has guided the proposal of Cu$_3$N$X_x$ as a promising candidate for the Dirac line node semimetal, where $X$ = Pd or Zn, and $x$ is a doping rate of the $X$ atom. It was found that doping a transition metal $X$ induces band inversion in Cu$_3$N without opening a band gap, giving rise to a Dirac line node. Independently, Cu$_3$NZn was proposed to be the topological nodal line semimetal by Rui Yu {\it et al.}~\cite{Yu15p036807}.  A large number of materials have been theoretically identified in this class of the nodal line semimetals, 
with the experimental realizations in graphite~\cite{Mikitik99p2147, Mikitik06p235112}, ZrSiS family~\cite{Schoop16p11696, Neupane16p201104, Hu16p016602, Hosen18p121103, Hosen18p1} and TiB$_2$ family~\cite{Zhang17p235116, Yi18p201107}, Mg$_3$Bi$_2$~\cite{Chang17p1711.09167}.

\subsection{$\mathbb{Z}_2$ monopole nodal line semimetals}

\begin{figure}[tb!]
\includegraphics[width=0.75\textwidth]{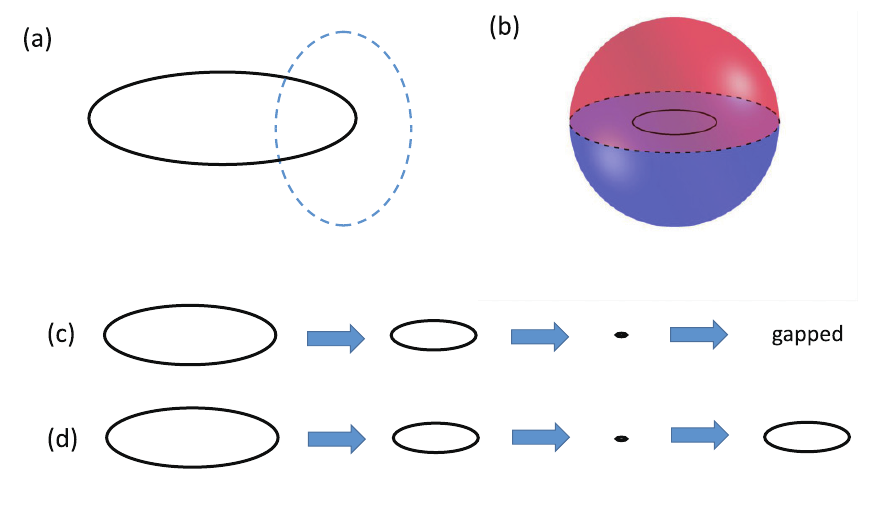}
\caption{ {\bf $\mathbb{Z}_2$ Berry  phase.}
 (a) A nontrivial $\mathbb{Z}_2$ Berry  phase of $\pi$ calculated on a loop in 3D BZ (dashed blue line) implies the presence of a nodal line (solid black line) passing through the loop. (b) A nontrivial $\mathbb{Z}_2$ monopole charge calculated on a 2D sphere implies the presence of a nodal line enclosed by the sphere. (c) Nodal line with nontrivial Berry phase $\pi$ and the zero monopole charge. It can be created (annihilated) alone as the bands are inverted (uninverted). (d) Nodal line with both nontrival Berry phase and the nonzero monopole charge. It can be reduced to a Dirac point but cannot be annihilated, but reverted into a nodal line. 
 {\it Adapted from Ref.~\cite{Fang15p081201}}
}
\label{fig:yk_fig3}
\end{figure}

In addition to the $\mathbb{Z}_2$ Berry phase, which characterizes a one-parameter family of Hamiltonians in 3D momentum space, a nodal line can also carry a $\mathbb{Z}_2$-quantized monopole charge when both $\mathcal{P}$ and $\mathcal{T}$ are present and spin-orbit coupling is vanishingly small. The $\mathbb{Z}_2$ monopole charge characterizes a two-parameter family of Hamiltonians in 3D momentum space ~\cite{Fang15p081201,Li17p247202,Song18p031069}. Unlike the nodal lines protected by the Berry phase, the nodal lines with a nontrivial monopole charge are stable in the sense that they can only removed by pairwise annihilation; an isolated monopole charge nodal line cannot be removed. This is similar to Weyl points discussed in Section \ref{sec:weyl} and is illustrated in Fig.~\ref{fig:yk_fig3}. Here, the $\mathbb{Z}_2$ monopole charge originates from the nontrivial second homotopy group of the classifying space of {\it real} Bloch states~\cite{Morimoto14p235127,Zhao17p056401,Fang15p081201}, in contrast with the integer monopole charge of {\it complex} Weyl fermions in systems with strong spin-orbit coupling.

\begin{figure}[tb!]
\includegraphics[width=0.8\textwidth]{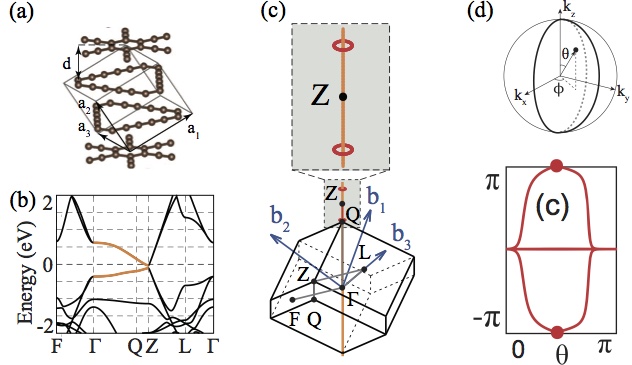}
\caption{{\bf $\mathbb{Z}_2$ monopole nodal line semimetal. }
DFT prediction for the $\mathbb{Z}_2$ monopole nodal line semimetal in an ABC stack of 2D graphdiyne. (a) Atomic structure, (b) electronic band structure. Orange lines indicate nodal lines formed between the two highest  occupied bands and the two lowest unoccupied bands. (c) $\mathbb{Z}_2$ nodal lines in the BZ. Red nodal lines are the $\mathbb{Z}_2$ monopole nodal lines, and the orange nodal line indicates the nodal line formed from the two highest occupied bands. (d) Wilson bands calculated on the sphere parameterized by polar coordinates shown in the top panel.  {\it Adapted from~\cite{Ahn18p106403}} }
\label{fig:yk_fig5}
\end{figure}

More specifically, the $\mathbb{Z}_2$ monopole charge manifests itself as a linking number which can be evaluated from the electronic band structure~\cite{Ahn18p106403}.  Ahn {\it et al.}~\cite{Ahn18p106403} have proved that the $\mathbb{Z}_2$ monopole charge is given by the linking number  
${\rm Lk}(\gamma,\tilde{\gamma}_{j})=\frac{1}{4\pi}\oint_{\gamma}d{\bf k} \times \oint_{\tilde{\gamma}_{j}} d{\bf p}\cdot \frac{{\bf k}-{\bf p}}{|{\bf k}-{\bf p}|^3}$, where $\gamma$ ($\tilde{\gamma}_{j}$) represents the nodal line formed from the conduction and valence (two top-most occupied bands). Furthermore, the $\mathbb{Z}_2$ monopole charge carried by a red nodal line can be represented by the second Stiefel-Whitney number $w_2$ defined as
\begin{align}
w_{2}
=\sum_{\tilde{\gamma}_{j}}{\rm Lk}(\gamma,\tilde{\gamma}_{j}),
\end{align}
which can be evaluated by the Wilson band calculations on a sphere enclosing the nodal line. Recently, the nodal line semimetal characterized by the $\mathbb{Z}_2$ monopole charge was identified in ABC-stacked graphdiyne shown in Fig.~\ref{fig:yk_fig5}(a)~\cite{Nomura18p054204, Ahn18p106403}. The Brillouin zone contains two nodal lines in the vicinity of the $Z$ point, which are linked by another nodal line formed between the highest occupied bands colored by orange in Figs.~\ref{fig:yk_fig5} (b) and (c). The corresponding linking number calculation gives a linking number of 1. The $\mathbb{Z}_2$ monopole charge is consistent with the liking number, which is evaluated from the Wilson bands on the sphere enclosing a nodal line, shown in Fig.~\ref{fig:yk_fig5} (d). 

\subsection{Mirror- and glide mirror-symmetry protected nodal line semimetals}

Crystals preserving a mirror ($\mathcal{M}$) or a glide-mirror ($\mathcal{G}$) symmetry can also host nodal lines on a mirror (glide)-invariant plane in momentum space. The bands with different mirror (glide-mirror) eigenvalues can cross without hybridizing on the two-dimensional invariant plane, resulting in degeneracies along one-dimensional lines. This mirror-protected nodal line is topological and carries a $\mathbb{Z}$ topological invariant $\zeta$ defined by 
\begin{equation}
\zeta = \frac{N_{A} - N_{B}}{2},
\end{equation}
where $N_{A}$ ($N_{B}$) is the number of occupied bands with mirror eigenvalue $+$ at two different points $A$ and $B$ on the mirror invariant plane. $|\zeta|$ dictates the number of nodal lines intersecting any line connecting the two points. In the absence of spin-orbit coupling a mirror-protected nodal line can be created by a band inversion of bands with distinct mirror eigenvalues. A line node of this kind constitutes an accidental band crossing and does not require a partner as in the case of $\mathbb{Z}_2$ monopole charge line nodes. Various materials in this class of topological semimetal have been theoretically identified,
including the experimentally confirmed materials, such as CaAgP \cite{Yamakage16p013708, Nayak18p1, Takane18p1} and PbTaSe$_2$ \cite{Sun17p077101, Okamoto16p123701, Xu17p064528, Takane18p1} families of materials.

\begin{figure}[tb!]
\includegraphics[width=0.8\textwidth]{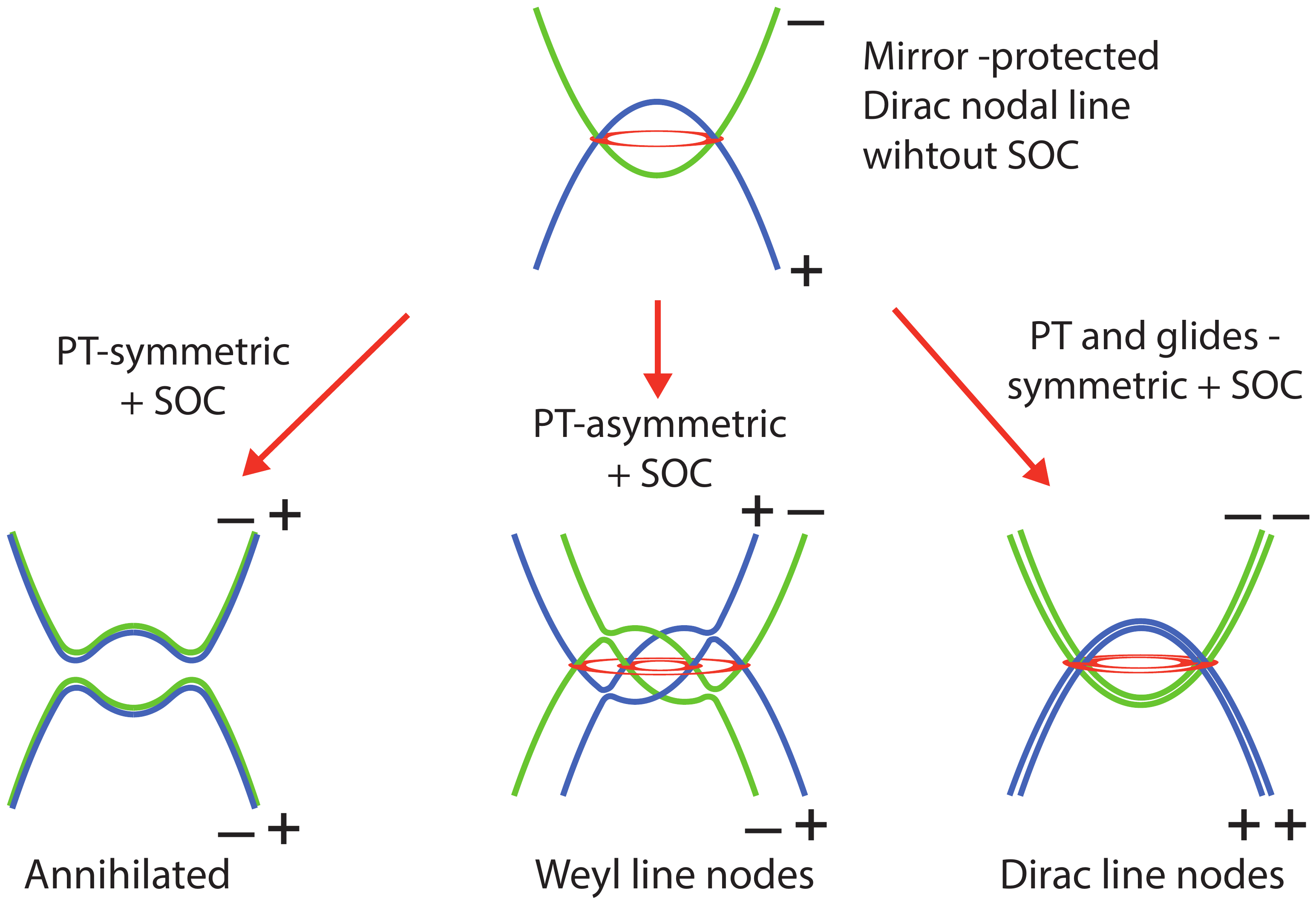}
\caption{{\bf Mirror protected nodal line semimetal. }
Schematic illustration of possible band evolution of mirror protected nodal line due to spin-orbit coupling. The mirror $-$ and $+$ bands are colored by green and blue, respectively.}
\label{fig:yk_fig6}
\end{figure}

In the presence of strong spin-orbit coupling, the mirror-protected nodal lines can be either annihilated or split into two twofold-degenerate Weyl line nodes--- a fate which is determined by the existence of the combined symmetry $\mathcal{PT}$, as shown in Fig.~\ref{fig:yk_fig6}. When $\mathcal{PT}$ is preserved, the Kramers pair $\psi$ and $\mathcal{PT}\psi$ should have opposite eigenvalues of the mirror operator.  For example, suppose $\psi_{\pm}$ is an eigenstate of $\mathcal{M}_z$ and thus satisfies $\mathcal{M}_z \psi_{\pm} = \pm i \psi_{\pm}$.  One then has $\mathcal{M}_z \mathcal{PT} \psi_+  = \mathcal{PT} \mathcal{M}_z \psi_+ = \mathcal{PT} (+i \psi_+) = -i \mathcal{PT}\psi_+$, which implies that $\mathcal{PT}\psi_+ \sim \psi_-$. Therefore, the twofold-degenerate Bloch states $\psi_{n \boldsymbol k}$ and $\mathcal{PT}\psi_{n \boldsymbol k}$ form a Kramers pair with the opposite mirror eigenvalues.  As a result, in the presence of spin-orbit coupling the energy bands forming the line node on the mirror invariant plane are allowed to hybridize and will anticross. This is shown in Fig.~\ref{fig:yk_fig6}.  In contrast, when either $\mathcal{P}$ or $\mathcal{T}$ is broken so that the Kramers pair is non-degenerate in energy, the bands with distinct mirror eigenvalues can cross each other forming a twofold-degenerate Weyl nodal line. Thus, the Dirac nodal line that appeared in the absence of spin-orbit coupling is split into two twofold-degenerate nodal lines when spin-orbit coupling is included in this $\mathcal{PT}$-asymmetric case.  Unlike the mirror-protected nodal lines, the nodal lines protected by nonsymmorphic space group symmetries, such as a glide-mirror or a screw, can survive even in the presence of spin-orbit coupling, retaining the fourfold-degenerate Dirac nodal lines~\cite{Fang15p081201,Wieder16p155108}. An orthorhombic perovskite iridate, such as AIrO$_3$, has been identified to host Dirac nodal line at the boundary of the BZ, protected by a screw and $\mathcal{PT}$~\cite{Chen15p1}.

\section{NEW FAMILIES OF TOPOLOGICAL SEMIMETALS \label{sec:generalized}  }

In the previous sections we have focused on TSMs defined by twofold- or fourfold-degenerate band touchings at isolated points or on lines in the Brillouin zone. In particular, we have discussed semimetallic systems which can be described by, or understood from, variations of the Dirac or Weyl equation. In this section, we turn to a discussion of more general classes of TSMs and focus on two classes in particular: stable symmetry-enforced band crossing points which cannot be described by Dirac or Weyl fermions, and ``triple-point'' fermion semimetals characterized by a symmetry-protected accidental crossing of three bands.

\subsection{Multifold band crossings in solids and low-energy fermions \label{ssec:new-fermions}}

As discussed in Section \ref{sec:dirac}, symmetry-enforced Dirac semimetals are defined by fourfold-degenerate band crossings located at high-symmetry points, with linear dispersion and vanishing Chern number. In solids with strong spin-orbit coupling and $\mathcal T$ symmetry, however, more general symmetry-enforced band crossings at points of high symmetry can occur, at which three, six, or even eight bands touch. The conditions for the existence of such nodal band degeneracies are determined by the structure of the crystal space group, which is required to include nonsymmorphic symmetry elements~\cite{Bradlyn16p5037}. The latter requirement implies that symmetry-protected degeneracy points of this kind are located at high-symmetry points on the BZ boundary. Note that when spin-orbit coupling is vanishingly small, threefold band crossings can also occur for symmorphic space groups~\cite{Manes12p155118}.

The presence of three-, six-, or eightfold band crossings gives rise to low-energy fermionic excitations which cannot be described by the Dirac or Weyl equations; instead, their dispersion is described by a more general Hamiltonian. The departure from the (low-energy) description of familiar relativistic free fermions (\ie Dirac and Weyl fermions) is a consequence of the less restrictive nature of the crystal symmetries governing condensed matter systems, which allow for the realization of generalized low-energy fermions. All of these TSMs characterized by ``multifold'' fermions (\ie three-, six-, and eightfold) are symmetry-enforced semimetals, as they fundamentally rely on the constraints originating from space group symmetry, possibly combined with $\mathcal T$ symmetry.

\begin{figure}[t]
\includegraphics[width=1.0\textwidth]{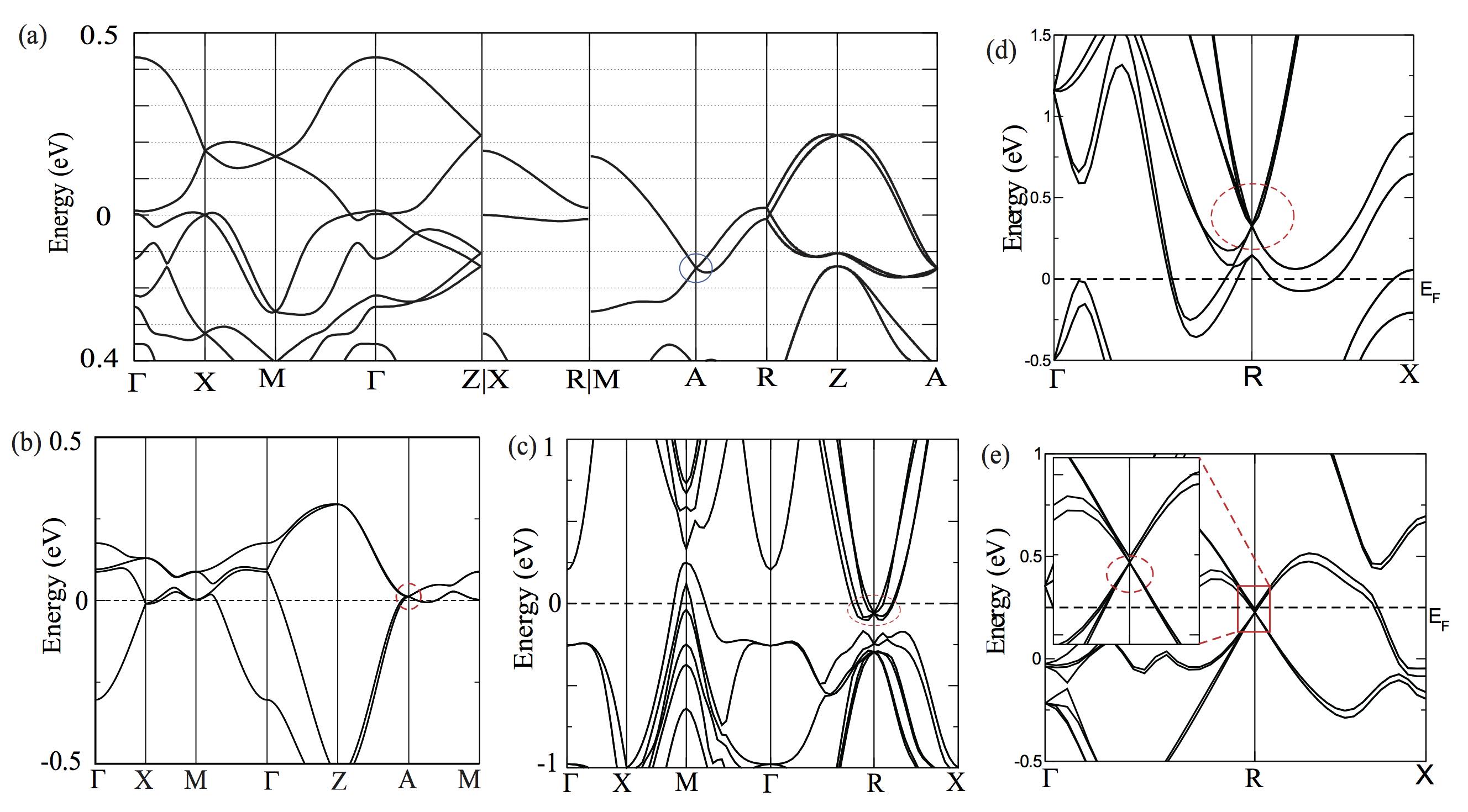}
\caption{{\bf Eightfold and sixfold fermions.} 
Examples of first-principles band structures showing eightfold and sixfold fermions at points of high symmetry on the Brioullin zone boundary. Eightfold fermions were predicted in (a) Ba$_2$AuO$_5$; (b) CuBi$_2$O$_4$; and (c) Ta$_3$Sb. Sixfold fermions were predicted in (d) MgPt and (e) Li$_2$Pd$_3$B.
{\it Adapted from the Refs.\cite{Wieder16p186402} and~\cite{Bradlyn16paaf5037}}}
\label{fig:8-6-fold}
\end{figure}

A first example was discussed by Wieder {\etal}~\cite{Wieder16p186402}, who studied and identified space groups which can protect eightfold-degenerate band crossings with linear dispersion. In particular, two tetragonal space groups, $\bf 130$ ($P4/ncc$) and $\bf 135$ ($P4_2/mbc$), were considered in detail, since in these space groups all energy levels at the $A$ point must be eightfold degenerate, \ie the $A$ point only admits eight-dimensional representations. Owing to the linear dispersion for all momenta $\delta \bk = \bk-\bk_0$ away from the touching point, such eightfold-degenerate points were referred to as ``double-Dirac'' points. Based on first-principles calculations, Wieder {\etal} predicted that the known material Bi$_2$AuO$_5$ in space group $\bf 130$ hosts a double-Dirac at the $A$ point, which sits close to the Fermi level~\cite{Wieder16p186402}. The calculated band structure for Bi$_2$AuO$_5$ is shown in Fig.~\ref{fig:8-6-fold} (a).

In subsequent work, Bradlyn {\etal} have considered the occurrence of space group symmetry-enforced multifold band crossings from a more general and comprehensive perspective~\cite{Bradlyn16p5037}. In particular, they exhaustively tabulated all space groups which, in combination with $\mathcal T$ symmetry, can stabilize three-, six-, or eightfold degenerate fermions, focusing on systems with spin-orbit coupling. Furthermore, the authors determined the general form of the Hamiltonian describing these ``unconventional fermions''. A series of material candidates was also proposed, including additional examples of materials in space groups $\bf 130$ and $\bf 135$ hosting eightfold fermions, such as the bismuth oxides CuBi$_2$O$_4$ and PdBi$_2$O$_4$ (space group $\bf 130$), as well as PdS (space group $\bf 135$). In CuBi$_2$O$_4$, shown in Fig.~\ref{fig:8-6-fold}(b), the eight-band fermion sits at the Fermi level, however, interaction effects may play an important role and render the material insulating~\cite{Sharma16p2936}. The material family $A_3B$ in space group $\bf 223$ (Pm3n), where $A$ is either Nb or Ta and $B$ is any group A-IV or A-V element in the $\beta$-tungsten structure A15, was identified as another candidate for hosting eight-band fermion, as evidenced by Ta$_3$Sb shown in Fig.~\ref{fig:8-6-fold} (c). Sixfold fermions were predicted in MgPt with space group $\bf 198$ ($P2_13$), shown in Fig.~\ref{fig:8-6-fold} (d), and in Li$_2$Pd$_3$B $\bf 212$ ($P4_332$), shown in Fig.~\ref{fig:8-6-fold} (e), among other examples.

The collection of proposed material candidates is encouraging, yet also showcases one of the remaining challenges for future research. In many examples---for instance those shown in  Figure~\ref{fig:8-6-fold}---the multifold fermion band crossings do not sit at the Fermi and (or) overlap in energy with generic Fermi surfaces. As a result, isolating the electronic excitations of the multifold fermions in real materials defines the most interesting---and perhaps most pressing---problem to attack in the field of symmetry-enforced semimetals.

\subsection{Topological fermions with higher spin\label{ssec:spin-j}}
 
The comprehensive classification of space group symmetry-protected band crossings obtained by Bradlyn {\etal} revealed that some of these band crossing points can be described by fermions with spin higher than $j = 1/2$. For instance, threefold degenerate point nodes give rise to excitations with spin $j=1$. Such higher spin systems form an interesting class of topological semimetals which deserves further discussion. 

\begin{figure}[t]
\includegraphics[width=1.0\textwidth]{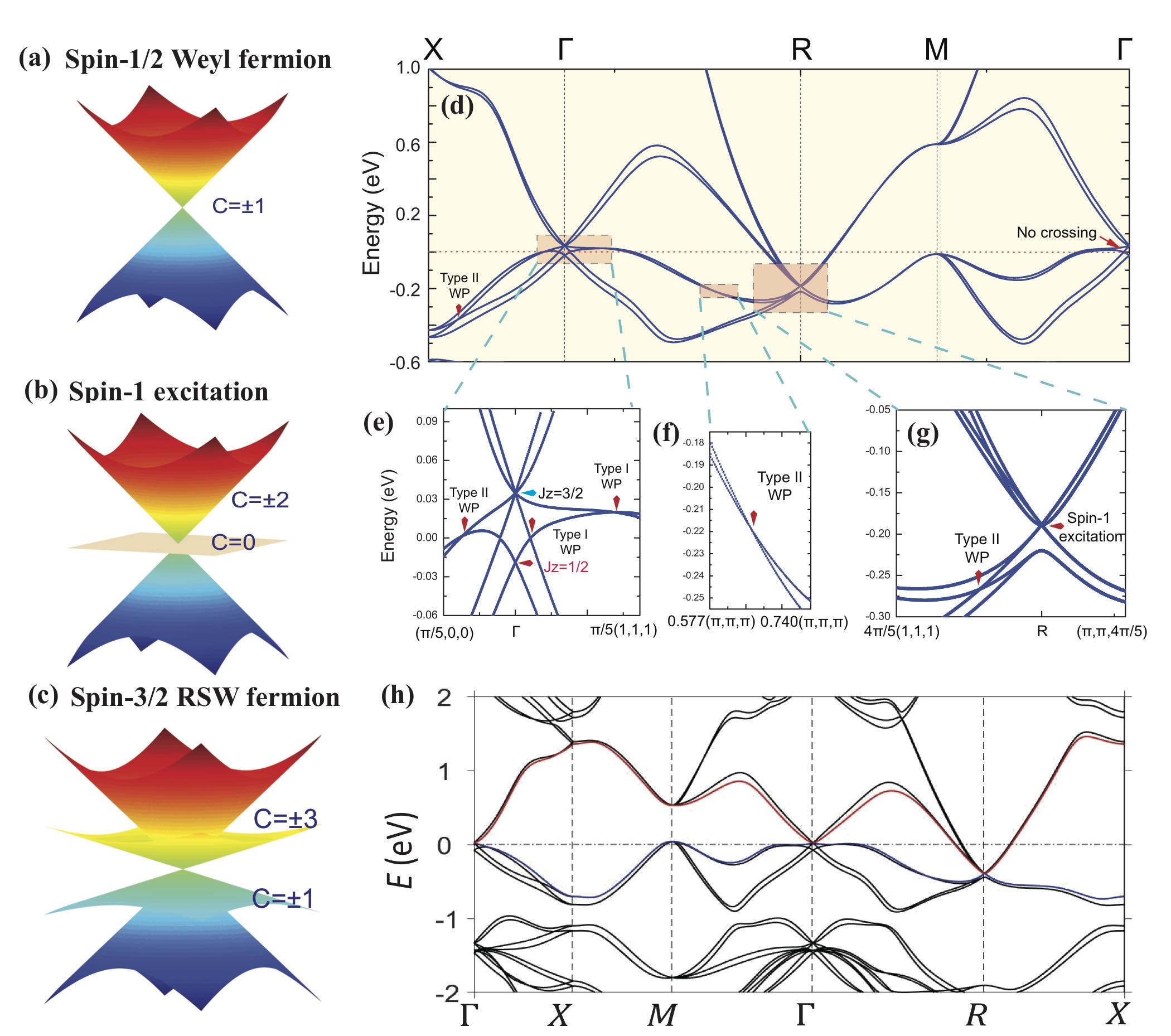}
\caption{{\bf Topological fermions with higher spin.} Panels (a)--(c) show the low-energy dispersions of $j=1/2$, $j=1$, and $j=3/2$ fermions, respectively, based on Eq.~\eqref{eq:H-spin-j}. The Chern numbers associated with these bands are indicated. 
Also shown are the bulk band structures of cubic silicides CoSi and RhSi in space group $\bf 198$ ($P2_13$) in the presence of spin-orbit coupling. 
Panel (a) shows the band structure of CoSi along high symmetry lines, which exhibits a fourfold-degenerate spin $j=3/2$ fermion at $\Gamma$ and a sixfold degenerate fermion at $R$. These are shown more clearly in the enlargements of panels (b) and (d), respectively. The sixfold fermion at $R$ can be described by a double spin $j=1$ fermion, as indicated in panel (d); the $j=3/2$ fermion at $\Gamma$ is marked by a blue arrow in (b). (Additional twofold Weyl points are present as well.) Panel (e) shows the band structure of RhSi with qualitatively the same features. For both materials the fourfold-degenerate fermion at $\Gamma$ lies at the Fermi energy, whereas the double spin $j=1$ fermion sits below the Fermi energy.
{\it Adapted from Ref.~\cite{Tang17p206402} and~\cite{Chang17p206401}.} }
\label{fig:spin-j}
\end{figure}

Fermions with higher spin can be viewed as generalizations of Weyl fermions. As discussed in Section \ref{sec:weyl}, the Hamiltonian of a Weyl fermion takes the form $H(\bm{k})=  v_{0} \delta\bm{k} \cdot \bm{\sigma}$, where $\delta\bm{k}$ is the momentum measured relative to the degeneracy point and $\bm{\sigma}$ are the Pauli matrices. As such, the Weyl Hamiltonian describes a low-energy gapless fermion with spin $j=1/2$. The generalization to higher spin is then straightforward and is achieved by replacing the Pauli spin matrices with the corresponding higher spin matrices. Specifically, the simplest Hamiltonian sufficient to capture the generic properties of higher spin fermions is given by 
\begin{equation}
H(\bm k)= v_0\delta \bm k\cdot\bm S,  \qquad   [S_i,S_j] = i\epsilon_{ijk}S_k,   \label{eq:H-spin-j}  
\end{equation}
where $\bm S = (S_x, S_y, S_z)$ are three spin matrices of a spin-$j$ fermion satisfying the canonical $SU(2)$ algebra. At $\delta \bm k =0 $ the bands are $2j+1$-fold degenerate, which shows that $j=1$ and $j=3/2$ fermions describe threefold and fourfold band crossings, respectively. The band dispersion of $j=1$ and $j=3/2$ fermions with Hamiltonian \eqref{eq:H-spin-j} is shown in Figs.~\ref{fig:spin-j}(b) and (c), which may be contrasted with the Weyl fermion shown in Fig.~\ref{fig:spin-j}(a).

Higher spin fermions are topological in the sense that they describe band crossings with net Berry monopole charge. The Berry flux associated with each band can be inferred from its rotation eigenvalues and gives $2j_z$ for $j_z=-j,-j+1,\ldots, j $.  For the $j=1$ and $j=3/2$ fermions described by \eqref{eq:H-spin-j} this implies that they carry monopole charge $C=\pm 2$ and $C= \pm 3 \pm 1=\pm 4$, respectively (see Fig.~\ref{fig:spin-j}). This may be compared to the Weyl fermion, which has monopole charge $C=\pm 1$. 

Even though Weyl fermions belong to the class of chiral topological fermions described by a Hamiltonian of the form of Eq.~\eqref{eq:H-spin-j}, it should be noted that Weyl fermions can occur at a generic point in the Brillouin zone, whereas fermions with higher spin can only occur at high-symmetry points. As mentioned in Section~\ref{ssec:new-fermions}, the threefold fermions with spin $j=1$ can only occur at high-symmetry points of the Brillouin zone boundary and require a nonsymmorphic space group. In contrast, the fourfold-degenerate band crossings with spin $j=3/2$ structure do not strictly rely on the nonsymmorphic nature of the space group and can occur for symmorphic space groups~\cite{Bradlyn16p5037}. The latter is not surprising, as it follows directly from the well-known fact that cubic double point groups can protect the degeneracy of a $j=3/2$ quartet. 

To showcase the space group classification of higher spin fermions, Bradlyn {\etal} proposed a number of material candidates to realize such semimetallic phases and supported these proposals with first-principles band structure calculations. Examples of materials predicted to exhibit threefold band crossings relatively close to the Fermi level are Pd$_3$Bi$_2$S$_2$ (space group $\bf 199$) and Ag$_3$Se$_2$Au (space group $\bf 214$). Furthermore, Ag$_3$Se$_2$Au was found to exhibit a fourfold band crossing at $\Gamma$, which is very close to the Fermi level. 

As part of this wave of proposals, topological fermions with higher spin were also predicted in a family of cubic silicide materials $A$Si and $A$Ge ($A = \text{Rh},\text{Co}$) in space group $\bf 198$ ($P2_13$)~\cite{Chang17p206401, Tang17p206402}. Figure~\ref{fig:spin-j} shows the bulk band structure of CoSi and RhSi reported in Refs.~\cite{Chang17p206401, Tang17p206402}, which exhibit a $j=3/2$ fermion at $\Gamma$ sitting at the Fermi energy and a sixfold-degenerate fermion at $R$. The latter is described by two copies of \eqref{eq:H-spin-j} and can thus be understood as a double spin $j=1$ fermion. The Berry monopole charge associated with these fermions (\ie the Chern numbers) exactly cancel ($4- 2\times 2=0$), as required by the Nielsen-Ninomiya theorem~\cite{Nielsen83p389} mentioned in Section~\ref{sec:weyl}. The semimetallic nature of CoSi and RhSi finds its origin in a combination of electron filling, space group symmetry, and $\mathcal T$ symmetry, which puts these materials in the class of filling-enforced semimetals~\cite{Watanabe15p14551, Wieder16p155108, Young17p186401}.

\subsection{Triple-point semimetals \label{ssec:triple}}

\begin{figure}[t]
\includegraphics[width=1.0\textwidth]{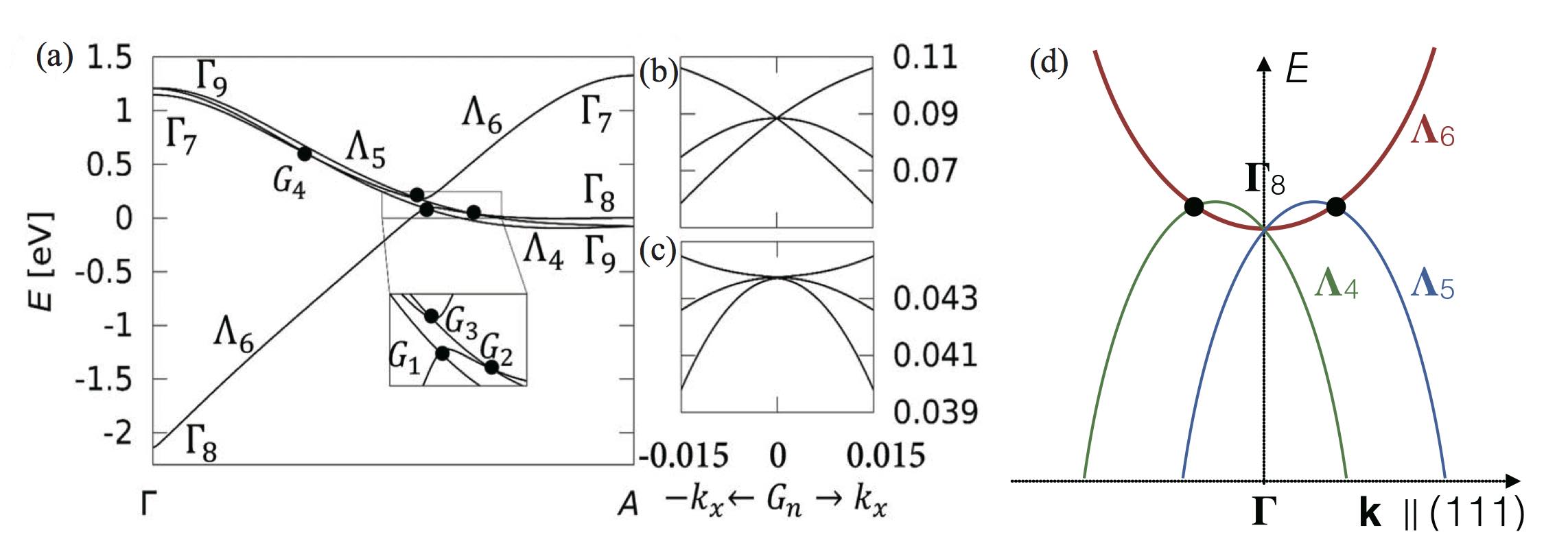}
\caption{{\bf Triple-point semimemtal.} Panel (a): Band structure of ZrTe along the $\Gamma$-A line. Bands are labeled by their double group representations corresponding to $D_{3h}$ at $\Gamma$ and A points and $C_{3v}$ on the $\Gamma$-A line. Panels (b) and (c) show band structures in the (100) direction with $k_{z}$ tuned to the TPs G$_1$ and G$_2$. Panel (d) shows the schematic band structure of unstrained HgTe near $\Gamma$. {\it Adapted from Ref.~\cite{Zhu16p031003}} }
\label{fig:triple}
\end{figure}

The multifold band crossings considered in Sections \ref{ssec:new-fermions} and \ref{ssec:spin-j} occur at special high-symmetry points in the Brillouin zone and are associated with higher dimensional space group representations of the corresponding little group. This is, however, not the only way in which stable threefold degenerate band crossings can arise in band structures. A symmetry-protected crossing of three energy bands can also occur at a (generic) point on a high-symmetry line, provided the little group of the high-symmetry line admits both one- and two-dimensional (double) group representations. This is true for threefold rotation axes with little group $C_{3v}$ and threefold band crossings of this kind have been identified as ``triple-point'' semimetals~\cite{Zhu16p031003, Weng16p241202, Chang17p1688, Wieder18p329}. 

The origin of triple-point semimetals can be traced back to a band inversion. As explained in the context of Dirac semimetals in Section~\ref{ssec:sym-invert}, band inversions can give rise to stable band crossings in the presence of symmetries, in particular rotation symmetry. To understand the origin of triple-point degeneracies one may consider the example of $j_z=\pm 1/2$ and $j_z=\pm 3/2$ bands inverting at $\Gamma$. Along the rotation axis with little group $C_{3v}$ the $j_z=\pm 1/2$ states must stick together, but the $j_z=\pm 3/2$ state generically split (except at $\mathcal T$-invariant points). This provides a simple picture of triple-point degeneracies and explains why they come in pairs. In addition, this example immediately reveals that further splitting the twofold degeneracy along the rotation axis by breaking appropriate symmetries results in a Weyl semimetal. 

The first proposals for observing this new type of TSM in real materials were focused on a family of two-element metals $AB$ with tungsten carbide (WC) structure, where $A=\{ \text{Zr}, \text{Nb}, \text{Mo}, \text{Ta}, \text{W} \}$ and $B=\{ \text{C}, \text{N}, \text{P}, \text{S}, \text{Te} \}$~\cite{Zhu16p031003, Weng16p241202, Chang17p1688}. The band structure of ZrTe is shown in Fig.~\ref{fig:triple}. Strained HgTe along the $(111)$ direction was suggested as another venue for these triple point band crossings~\cite{Zhu16p031003}. Since HgTe is structurally related to the half-Heusler compounds (the latter crystallize in a stuffed zincblende structure) and have similar electronic structure~\cite{Chadov10p541}, subsequent work expanded the search for triple-point semimetals to this material class~\cite{Yang17p136401}. The layered trigonal material PtBi$_2$ was proposed as yet another possible venue~\cite{Gao18p3249}. Remarkably, angle-resolved photoemission measurements performed on MoP~\cite{Lv17p627} and WC~\cite{Ma18p349} have since led to a first report of experimental observation of triple-point fermions.

\section{STRATEGIES FOR MATERIAL PREDICTION AND DESIGN \label{sec:prediction}  }

In this final section we discuss recent developments in the design and implementation of systematic materials search strategies, geared towards the prediction novel TSMs. The proposals for material realizations discussed in this review originated from the attempt to find examples of a specific type of TSM. In these targeted searches, identification of materials or material classes was predicated on the specific symmetry requirements and electrochemical parameters (\eg orbital composition, spin-orbit coupling) associated with a particular TSM, and in many cases a great deal of physical or chemical intuition was involved. As successful as such approaches based on expertise and experience have proven to be, the goal to perform more systematic and algorithmic searches of known synthesized materials has motivated attempts to put the classification and diagnosis of topological electronic band structures on a new footing. 

These attempts have led to a set of recent ideas centered around the notion of symmetry indicators~\cite{Bradlyn17p298, Po17p50, Kruthoff17p041069, Song18p3530, Song18p031069}. The general philosophy underlying these ideas is to use the symmetry quantum numbers characterizing a band structure (\eg space group representations at high-symmetry points), as well as the constraints which these must satisfy (\eg compatibility relations), and infer from this symmetry data the topological characteristics of the band structure, \ie the topological data. Specific cases where such an approach applies were known, as for instance the inversion-eigenvalue formula for detecting Weyl points~\cite{Hughes11p245132}, and the methods relying on symmetry indicators promote these specific examples to a comprehensive and systematic framework. This is particularly powerful in the context of first-principles calculations, in which the symmetry data is more readily accessible than the topological data. 

Remarkably, a series of very recent works have employed this set of ideas to systematically search large material databases for previously unrecognized topological materials, both insulators and semimetals~\cite{Zhang18p08756,Vergniory18p10271,Tang18p09744}. Based on high-throughout computational schemes, thousands of new topological materials were identified, constituting an encouraging step forward in the quest for new and better topological materials.

\section{SUMMARY AND OUTLOOK}

In this review we have given an overview of some recent developments in the field of TSMs. In addition to the Dirac and Weyl semimetal phases, which provide condensed matter realizations of Dirac and Weyl fermions, our survey has focused on some more recent examples of TSMs, such as nodal line semimetals, multifold fermion semimetals, and triple-point fermion semimetals. In a broad sense, two concepts were shown to be central to the understanding of TSMs: the notion of symmetry-enforced band crossings and the notion of a band inversion. The latter establishes a direct connection to other classes of topological materials, such as topological insulators and topological crystalline insulators. The examples of proposed material realizations discussed in this review showcase the importance of first-principles band structure calculations for the prediction of new topological materials, in particular TSMs.

A number of important and exciting challenges lie ahead. Building on the encouraging experimental observation of some of these TSMs, future efforts will concentrate on identifying ``ideal'' versions of TSMs in real materials. TSMs can be called ideal when the low-energy gapless fermions sit at the Fermi energy and do not overlap in energy with other (non-topological) band structure features. Another direction for the future is to explore the properties of TSMs in more detail, such as transport and optical properties. The ability to control and manipulate these properties will be a prerequisite for using TSMs as a resource for quantum devices.

\section*{DISCLOSURE STATEMENT}

If the authors have noting to disclose, the following statement will be used: The authors are not aware of any affiliations, memberships, funding, or financial holdings that
might be perceived as affecting the objectivity of this review. 

\section*{ACKNOWLEDGMENTS}
Our own research in the general area of topological materials has greatly benefited from current and past collaborations with many distinguished colleagues. We feel particularly indebted to Liang Fu, Charles Kane, Eugene Mele, Benjamin Wieder, Steven Young, Saad Zaheer. We owe special thanks to Benjamin Wieder and Eugene Mele for helpful correspondence and a careful reading of a draft version of the manuscript. 
JWFV was supported by the National Science Foundation MRSEC
Program, under grant DMR-1720530. YK was supported from Institute for Basic Science (IBS-R011-D1) and a National Research Foundation of Korea (NRF) grant (NRF-2017R1C1B5018169). AMR was supported by the Department of Energy, Office of Basic Energy Sciences, under grant DE-FG02-07ER46431.

\bibliographystyle{ar-style3}

\end{document}